\documentclass[a4paper,11pt]{article}

\usepackage[utf8]{inputenc}
\usepackage[english]{babel}
\usepackage[myheadings]{fullpage}
\usepackage[T1]{fontenc}
\usepackage{subfig}
\usepackage{fancyhdr}
\usepackage{graphicx}
\usepackage{sectsty}
\usepackage{url}
\usepackage[usenames, dvipsnames, table, xcdraw]{xcolor}
\usepackage{amsmath}
\usepackage{amssymb}
\usepackage{calrsfs}
\usepackage{bbold}
\usepackage{dsfont}
\usepackage{esint}
\usepackage{indentfirst}
\usepackage{xfrac}
\usepackage{bigints}
\usepackage{slashed}
\usepackage{breqn}
\usepackage{autobreak}
\usepackage{setspace}
\usepackage[colorlinks]{hyperref}
\usepackage{microtype}
\usepackage{quoting}
\usepackage{bigints}
\usepackage{float}
\usepackage{array}
\usepackage[export]{adjustbox}
\usepackage[section]{placeins}
\usepackage[capitalise]{cleveref}
\usepackage{enumitem}
\usepackage{scalefnt}
\usepackage{booktabs}
\usepackage{caption}
\usepackage{makecell}
\usepackage{authblk}

\usepackage[normalem]{ulem}

\usepackage{cite}

\usepackage{xspace}

\usepackage{colortbl}
\definecolor{myblue}{rgb}{0.8,0.85,1}
\definecolor{light-gray}{gray}{0.95}

\hypersetup{colorlinks=true,
citecolor=ForestGreen,
anchorcolor=CornflowerBlue,
linkcolor=NavyBlue,
urlcolor=Mahogany,
linkbordercolor={1 1 0}}

\makeatletter
\newcommand{\thickhline}{\noalign {\ifnum 0=`}\fi \hrule height 1pt
\futurelet \reserved@a \@xhline}
\newcolumntype{"}{@{\hskip\tabcolsep\vrule width 1pt\hskip\tabcolsep}}
\makeatother

\allowdisplaybreaks

\def\beq {\begin{equation}}
\def\eeq {\end{equation}}
\def\bea {\begin{eqnarray}}
\def\eea {\end{eqnarray}}

\def\nn {\nonumber}

\AtBeginDocument{}

\title{\LARGE {\bf \sffamily \boldmath Probing the Singularities of the \\[2mm]
Landau-Gauge Gluon and Ghost Propagators \\[2mm]
with Rational Approximants}}

\author[a]{Diogo Boito}
\author[a]{Attilio Cucchieri}
\author[a,b]{Cristiane Y. London\thanks{cristiane.london@usp.br (corresponding author)\vspace{0.3cm}}}
\author[a]{Tereza Mendes}

\affil[a]{\it Instituto de F\'isica de S\~ao Carlos, Universidade de S\~ao Paulo, CP 369, 13560-970, S\~ao Carlos, SP, Brazil\vspace{0.2cm}}

\affil[b]{\it Grup de F\'isica Te\`orica, Departament de F\'isica, Universitat Aut\`onoma de Barcelona, and Institut de F\'isica d'Altes Energies (IFAE), The Barcelona Institute of Science and Technology (BIST), Campus UAB, E-08193 Bellaterra (Barcelona), Spain\vspace{0.3cm}}

\date{}

\begin{document}

\begin{flushright}
{\small \today}
\end{flushright}

\vspace*{-0.7cm}
\begingroup
\let\newpage\relax
\maketitle
\endgroup

\vspace*{-1.0cm}
\begin{abstract}
\noindent
We employ Pad\'e approximants in the study of the analytic structure of the
four-dimensional $SU(2)$ Landau-gauge gluon and ghost propagators in the
infrared regime.
The approximants, which are model independent, serve as
fitting functions for the lattice data. We carefully propagate the
uncertainties due to the fitting procedure, taking into account all possible
correlations.
For the gluon-propagator data, we confirm the presence of a pair of complex
poles at $p_{\rm pole}^2 = \left[(-0.37 \,\pm\, 0.05_{\rm stat}\,\pm\, 0.08_{\rm sys})
\pm i\,(0.66\, \pm\, 0.03_{\rm stat}\, \pm\, 0.02_{\rm sys})\right]\,
\mathrm{GeV}^2$, where the first error is statistical and the second systematic.
The existence of this pair of complex poles, already hinted upon in previous works, is thus put onto
a firmer basis, thanks to the model independence and to the careful
error propagation of our analysis.
For the ghost propagator, the Pad\'es indicate the existence of a single pole
at $p^2 = 0$, as expected.
In this case, our results also show evidence of a branch cut along
the negative real axis of $p^2$. This is corroborated with another type of
approximant, the D-Log Pad\'es, which are better suited to studying functions
with a branch cut and are applied here for the first time in this
context.
Due to particular features and limited statistics of the gluon-propagator
data, our analysis is inconclusive regarding the presence of a branch cut in
the gluon case.
\end{abstract}

\thispagestyle{empty}

\clearpage
\vspace*{0.0cm}
\tableofcontents

\setcounter{page}{1}

\vspace{1cm}

\section{Introduction}

The low-momentum behavior of Green functions in Yang-Mills theory for
different gauges has been an important subject of study for the past 40
years or so, mainly in connection with the understanding of the
color-confinement property
of QCD~\cite{Alkofer:2000wg,Greensite:2011zz,Vandersickel:2012tz,Pasechnik:2021ncb}.
The topic is treated both by analytic and numerical methods, all of which
involve significant technical challenges.
In particular, most studies have
focused on the Landau-gauge case, giving rise to several theoretical advances
and proposals, with lively debate between researchers using different
approaches~\cite{Cucchieri:2009xxr}.

Concerning lattice simulations of Green functions, the main numerical
difficulty is somewhat unusual.
Indeed, in typical lattice-QCD
applications, one just needs to consider a large enough
discretized volume to ensure a physical lattice side $L$ a few times greater
than the relevant hadronic scale ($\approx 1~\text{fm}$). This stems
from the fact that finite-size corrections affecting hadronic observables are
suppressed by a factor of $\exp{(-m_{\pi}L)}$~\cite{Aoki:2021kgd}, where
$m_{\pi}$ is the lattice pion mass.
The effort, then, is to go to very small lattice spacing
($\lesssim 0.1~\text{fm}$), in order to avoid discretization errors.
In studies of the infrared (IR) region, however, the situation is different.
This happens because the IR limit, which corresponds to small momenta,
requires
a large lattice side $L$. In fact, the smallest nonzero momentum that can be
represented on a lattice is $\approx 2\pi/L$.
As it turns out, for the Landau-gauge gluon and ghost propagators, finite-size effects
are particularly severe~\cite{Cucchieri:2003di,Bogolubsky:2007ud,Cucchieri:2007md,
Sternbeck:2007ug,Boucaud:2011ug,Oliveira:2012eh,Bornyakov:2013pha,Cucchieri:2016qyc}
(see also Ref.~\cite{Maas:2011se} and references therein),
implying that one needs data at momenta $p$ well below $300~\text{MeV}$ in
order to obtain a clear picture of the IR region.
The low-momentum behavior of the gluon and ghost propagators is a central prediction
of different theoretical frameworks --- or scenarios --- for understanding
the confinement mechanism. Early on, the lack of control over finite-size
effects in simulations has obscured the true qualitative behavior of
these propagators in the deep IR region. This has been settled by later
studies~\cite{Cucchieri:2009xxr}.
Nevertheless,
present lattice data still accommodate descriptions based on rather different
analytic predictions~\cite{Fischer:2006ub,Boucaud:2011ug,Aguilar:2015bud,Huber:2018ned,
Pelaez:2021tpq,Dupuis:2020fhh}.
On the contrary, violation of reflection positivity in the real-space
gluon propagator, which may be claimed to be a signal of color
confinement~\cite{Alkofer:2000wg,Greensite:2011zz,Vandersickel:2012tz,Pasechnik:2021ncb},
has been a directly observed feature since early simulations~\cite{Mandula:1987rh,
Cucchieri:2004mf,Bowman:2007du}.
Recently, the focus has switched from testing specific analytic predictions
to the investigation
of the analytic structure of the IR gluon and ghost propagators, more or less
independently of the various scenarios. In particular, the main effort has
been in identifying the dominant singularities (poles, branch cuts) in the
complex plane. This usually involves either direct analytic studies using
complex momenta~\cite{Strauss:2012dg,Strauss:2012zz,Fischer:2020xnb,
Horak:2021pfr,Horak:2021syv,Horak:2022myj} or an analytic continuation of the
Euclidean Green functions by different methods \cite{Dudal:2019gvn,Li:2019hyv,
Binosi:2019ecz,Hayashi:2021nnj,Hayashi:2021jju,Li:2021wol,Lechien:2022ieg},
guided by general properties of the Källén-Lehmann spectral density.

The use of rational (or Pad\'e) approximants provides a model-independent
route to attack this problem.
Recently, Pad\'e approximants were used to fit $SU(3)$
propagators~\cite{Falcao:2020vyr,Oliveira:2021job} by Falc\~ao, Oliveira
and Silva. Here, we follow this approach to study
the analytic structure of the four-dimensional $SU(2)$ Landau-gauge
gluon and ghost propagators in the IR regime. More specifically, we
carry out a consistent and systematic analysis using Pad\'e approximants (PAs)
(and other rational approximants) as fitting functions to
the lattice data
from Refs.~\cite{Cucchieri:2007md,
Cucchieri:2007rg,Cucchieri:2008fc,Cucchieri:2009xxr,Cucchieri:2011ig,Cucchieri:2016jwg}.
Let us recall that a Pad\'e approximant $P^M_N(z)$
is the ratio of two polynomials of degree $M$
(in the numerator) and $N$ (in the denominator).
Pad\'e theory~\cite{Baker1975essentials,Baker1996pade} then
provides us with a systematic procedure to reconstruct a given function,
using the same information as its truncated Taylor expansion:
the derivatives of the function at a given point in the complex plane,
often taken to be the origin.
However, note that, compared with the Taylor series --- and
provided sufficient information about the function is available --- Pad\'e
approximants are much more powerful. Indeed, they 1) have an extended
range of validity, 2) are capable of reproducing the singularities of
the original function (poles and residues), and 3) can mimic the
existence of branch cuts in the
complex plane. Moreover, in certain situations, theorems
guarantee the convergence of sequences of approximants to the original
function in a given limit, except at the singularities where the function is not well
defined. Nevertheless, even when theorems are not available,
Pad\'e approximants have been shown in practice to be extremely useful as
well~\cite{MasjuanQueralt:2010hav}. We refer to the procedure of building
PAs from the Taylor expansion of a function at a given point
as ``genuine Pad\'es''. These have found many applications in particle
physics (for a few recent ones see
Refs.~\cite{Masjuan:2007ay,Hoang:2008qy,Masjuan:2009wy,Masjuan:2013jha,Caprini:2016uxy,
Aubin:2012cc,Aubin:2012me,Boito:2018rwt,Boito:2021scm,Ananthanarayan:2020umo}).

Per contra, in the present work, we employ a variant of the above
procedure, using rational approximants as fitting functions to
(lattice) data, taking advantage of the extended range of validity of
the Pad\'e approximants compared to the Taylor expansion.
This procedure departs from the genuine Pad\'es but it has been successfully
used in several applications to experimental
data~\cite{Masjuan:2008fv,Masjuan:2012wy,
Escribano:2013kba,Masjuan:2015cjl,Masjuan:2017tvw,VonDetten:2021rax}, as well
as to $SU(3)$ lattice data for Landau-gauge propagators~\cite{Falcao:2020vyr,
Oliveira:2021job}. In fact, using rational approximants as fitting
functions can be a powerful method, especially
in situations where the precise theoretical description of the data is not
feasible or is model dependent. In particular, the use of sequences
of approximants provides a model-independent way of extracting
crucial information, such as resonance
poles in scattering amplitudes or the Taylor expansion of hadronic form factors.
Here, we aim at extracting information about the
analytic structure of the Landau-gauge gluon and ghost propagators.

We point out that a popular confinement scenario proposes for the IR gluon
propagator a form that can be cast as a Pad\'e approximant.
Indeed, the Gribov-like~\cite{GRIBOV19781} (or Stingl-like~\cite{Stingl:1985hx,Stingl:1994nk})
function $f_1(p^2)$
of Ref.~\cite{Cucchieri:2011ig} belongs to the Pad\'e sequence
$P_{N+1}^N(p^2)$ (for $N=1$).
Here we do not assume any specific rigid
form for the propagators, but rather we explore their properties by
a systematic investigation of different Pad\'e sequences,
in a model-independent way.
As said above, a first analysis of the singularities of
the $SU(3)$ Landau-gauge gluon and ghost propagators
using rational approximants was presented in
Refs.~\cite{Falcao:2020vyr,Oliveira:2021job}.
In their work, the authors found evidence for the existence
of a pair of complex poles in the gluon propagator and of
a simple pole at zero momentum in the ghost-propagator data.
At the same time, their results seem to support the presence of a branch
cut --- along the negative real axis of the Euclidean $p^2$ momenta --- for both
propagators. Similar outcomes are obtained in Ref.~\cite{Binosi:2019ecz}
for gluon and ghost propagators, both in Landau gauge and in linear
covariant gauge, using a related approach.

Although our analysis and that of Refs.~\cite{Falcao:2020vyr,Oliveira:2021job}
are very similar in spirit, there are a few important differences.
Firstly, we consider $SU(2)$ lattice-gauge-theory
data~\cite{Cucchieri:2007md,Cucchieri:2007rg,Cucchieri:2008fc,Cucchieri:2009xxr,
Cucchieri:2011ig}, whereas the authors of Refs.~\cite{Falcao:2020vyr,Oliveira:2021job}
analyzed gluon and ghost propagators in the $SU(3)$ case.
(We recall, however, that the propagators are very similar for
these two gauge groups in the IR limit, not only qualitatively but also
quantitatively \cite{Cucchieri:2007zm,Cucchieri:2007ji,Cucchieri:2007bq}.)
Secondly, we perform a careful analysis of the uncertainties
involved in the fitting procedure.
Note that this is especially relevant for results from PAs with many
parameters, since these are likely affected by uncontrolled errors, thus
compromising the reliability of the final results for the analytic structure
of the propagators in the complex plane.
Indeed, since we use Pad\'e approximants as fitting functions,
a $\chi^2$ fit procedure is clearly required, and
the uncertainties in the parameters of the fit must be carefully
examined.
Moreover, the errors have to be propagated to all
related results, including the position of zeros and poles
in the complex plane.
Consequently, these uncertainties, which ultimately reflect the
information available in the data set, restrict the number of parameters
that can be reliably obtained in a given fit, thus imposing limits on
the order of the approximants $P_N^M(z)$ that can be employed in practice.
In particular, we will see that, for the data sets we have, our analysis
is limited to PAs of relatively low order (up to 10 parameters).
In the present study, all uncertainties in the fits to data are carefully
calculated (with more than one method), including all correlations when
necessary, and
the limitations they impose on our analysis are discussed in detail.

This paper is organized as follows. In Sec.~\ref{sec:pade} we give an
overview of Pad\'e theory with emphasis on the present application.
In Sec.~\ref{sec:examples} we discuss the use
of the approximants to analyze a toy data set,
in order to check the applicability of our approach.
In Sec.~\ref{sec:data} the details of the lattice simulation and of
the lattice data considered here are briefly described.
Our results are given in Sec.~\ref{sec:res}, with Sec.~\ref{sec:gluonprop}
devoted to
the application of PAs to the Landau-gauge gluon propagator, and
Sec.~\ref{sec:ghostprop} to applications to the ghost propagator
using PAs and partial PAs, obtained by imposing the existence
of the pole at zero momentum.
For the ghost propagator we also consider in Sec.~\ref{sec:dlog}
the use of D-Log Pad\'e approximants~\cite{Baker1996pade,Boito:2018rwt,Boito:2021scm},
a variant of the method better suited for the application to functions
that present a branch cut.
Our conclusions are given in Sec.~\ref{sec:conclusions}.
Some technical details on fits for highly-correlated data, and in
particular a short description of the so-called ``diagonal fits'',
are reported in Appendix~\ref{sec:AppCorrData}.
In Appendix~\ref{app:Precision}, we briefly discuss the issue of numerical
precision when using higher-order Padé approximants.

\section{Rational approximants}\label{sec:pade}

In this section we introduce the most important concepts about Pad\'e approximants, with emphasis on the problem we have at hand. A much more comprehensive discussion, including the demonstration of theorems, can be found in the works by Baker and Graves-Morris~\cite{Baker1975essentials,Baker1996pade} (see also
Ref.~\cite{MasjuanQueralt:2010hav} for several applications in particle physics).

The Pad\'e approximant, $P_N^M (z)$, is defined as the ratio of the polynomials of order $M$ and $N$, $Q_M(z)$ and $R_N(z)$, respectively, with $R_N(0) = 1$:
\begin{equation}
P_N^M (z) = \dfrac{Q_M(z)}{R_N(z)} = \dfrac{a_0 + a_1 \, z + \cdots + a_M \, z^M}{1 + b_1 \, z + \cdots + b_N \, z^N} \; .
\label{eq:pa}
\end{equation}
If the Taylor series of a function $f(z)$ is known, the canonical procedure to build PAs to this function is to determine the coefficients $a_k$ and $b_k$ by matching the expansion of $P^M_N(z)$ to the first $M+N+1$ coefficients of the Taylor expansion of $f(z)$.

In Pad\'e theory, convergence theorems are available for analytic and single-valued functions with multipoles or even essential singularities~\cite{Baker1996pade}. A class of functions that play a prominent role are the Stieltjes functions, which can be written in integral form as
\begin{equation}
f(z) = \int\displaylimits_0^\infty \dfrac{\mathrm{d}\phi(u)}{1+zu} \; ,
\end{equation}
where $\phi(u)$ is a (non-negative) measure on $[0,\infty)$.
Indeed, for $f(z)$, the Pad\'e sequences $P_N^{N+k}(z)$, with $k \geq -1$,
converge to Stieltjes functions, with some interesting properties.
One of them is that the poles of these Pad\'es are always simple and are located on the negative real axis of $z$ with positive residues~\cite{Baker1996pade}.
The connection with the present work is obvious from the K\"all\'en-Lehmann
representation of propagators, in terms of positive-definite spectral
functions.\footnote{We recall, however, that --- as stressed in
the Introduction --- positivity-violation is considered well established
for the Landau-gauge gluon propagator.}
As we will see, the Pad\'es to the gluon propagator $D(p^2)$ have poles
with an imaginary part incompatible with zero, which is in contradiction
with the usual K\"all\'en-Lehmann representation for $D(p^2)$~\cite{PhysRevD.99.074001,
Kondo:2019rpa}.
Following Ref.~\cite{Horak:2022myj}, this would also imply the existence
of several branch cuts for the gluon and ghost propagators, and the
consequent violation of the K\"all\'en-Lehmann representation
(see also Ref.~\cite{Horak:2021pfr}) for the ghost propagator $G(p^2)$.
(For general discussions of the analytic structures of these propagators
see also Refs.~\cite{Lowdon:2017gpp,Hayashi:2021nnj,Hayashi:2021jju}.)
This suggests that an integral representation for the gluon and
ghost propagators is only possible if the function is not Stieltjes,
i.e.\ if one considers some modified spectral representation~\cite{Binosi:2019ecz,
Lechien:2022ieg,Horak:2022myj}.

Another important result from Pad\'e theory that will guide our work is
related to Pommerenke's theorem, which states that a sequence of approximants
$P^N_{N+k}(z)$ --- built for a meromorphic function
$f(z)$ --- converges in any compact set of the complex plane
[except for a set of zero area containing the poles of $f(z)$].
Indeed, as $N$ increases, the poles of $f(z)$ are well
reproduced by the PAs and tend to be stable, i.e.\ they remain
unchanged (or almost unchanged) when $N$ is increased.
At the same time, extraneous poles can also appear in the
PAs, but they either go away when the order is increased or they appear
in combination with a nearby zero, which leads to a very small residue,
partially cancelling the effects of the pole.
The latter case corresponds to the so-called
{\it Froissart doublets}, which do not represent genuine poles of $f(z)$.
Note that the appearance of transient Froissart doublets may
``delay'' the convergence of a Pad\'e sequence to a function,
since they effectively reduce the order of the approximant~\cite{Boito:2018rwt,Boito:2021scm}.
Nevertheless, Pad\'e approximants that have doublets can
still be used to approximate the function away from these singularities.

Functions with branch points and cuts can also be approximated by PAs,
although for such functions the expected convergence
is mainly motivated by experience and not by theorems.
In this case, the
approximants will mimic a cut of a given function by
accumulating poles (interleaved with zeros) along the cut in the complex
plane~\cite{Baker1996pade,MasjuanQueralt:2010hav,Costin:2021bay}. Since the
gluon and ghost propagators are expected to have a
cut~\cite{Falcao:2020vyr,Oliveira:2021job,Horak:2022myj},
this will be of relevance here.

As said in the Introduction, we use the PAs as
fitting functions to describe (lattice) data sets.
The advantages of this procedure are 1) its model independence,
2) the fact that it can be applied in a systematic way,
and 3) the connection with Pad\'e theory (although here no theorems are
available).
This type of application is quite common in other particle-physics
problems for which the theoretical description of the data is
model dependent or
incomplete~\cite{Masjuan:2008fv,Masjuan:2012wy,Masjuan:2013jha,Masjuan:2015cjl,
Masjuan:2017tvw,VonDetten:2021rax,Caprini:2016uxy}.
An application that is particularly close to ours is the
extraction of resonance poles, using the PAs as fitting functions
to decay, scattering or form-factor data~\cite{Masjuan:2013jha,Caprini:2016uxy,VonDetten:2021rax}.
Indeed, it has been shown that the use of PAs as fitting functions
is a reliable and model-independent method to determine resonance pole
positions from fits to data sets. The precision of this procedure is
obviously limited by the quality of the data set: with larger errors, less
information is available and, eventually,
adding more parameters --- i.e.\ increasing the order of the PA in the
sequence --- is no longer an advantage, since the errors of the
parameters (and of the pole positions) increase considerably.
In some applications, it has been found that the maximum number of parameters
that can be meaningfully extracted from a data set was 6 or 7, which
limits the PAs of the $P_N^N(z)$ sequence to $N=3$, for
instance~\cite{Masjuan:2013jha}.
We will show that, with our lattice data for the gluon propagator,
the situation is very similar: it hardly makes sense to have more than
6 or 7 parameters in the PAs.
Likewise, for the ghost data, we were able to consider at most 7 or 8
parameters in the Pad\'e approximants.
Of course, with increased statistics, one expects that more parameters
would be
allowed and that fits with higher-order PAs would be meaningful as well.

\section{Conceptual examples}\label{sec:examples}

Before applying the procedure to the real lattice data, it is useful to
test the method in a fully controlled setting,
i.e.\ applied to a function whose analytic structure is known. This is
especially important here, since the application of PAs as fitting
functions is beyond the reach of the theorems of Pad\'e theory.
We start by building genuine PAs, from the exact knowledge of the Taylor
series of the chosen function. Then, we generate
a toy data set, following the main features of the lattice
data we will analyze in the next sections.
Similar tests with toy data sets were also carried out in Ref.~\cite{Falcao:2020vyr}.

For the present analysis, we choose a function that has as many features
as possible --- poles, a branch cut, as well as
zeros --- in order to mimic a sufficiently realistic scenario.
To this end, we employ the function
\begin{equation}
g(z) = 1.15 \, \log{(z + 2)} \,\, \dfrac{z + 2.508}{z^2 + 0.768^2\,z + 0.72^2} \; ,
\label{eq:logpole}
\end{equation}
which roughly resembles the lattice data for the gluon propagator.
In fact, this function has a pair of complex poles at
$z_0 = -0.295 \pm 0.657 \, i$, two zeros (located at $z = -1$ and
$z = -2.508$) and a branch cut from $(-\infty, -2]$, similarly to the
results presented in Ref.~\cite{Cucchieri:2011ig}.
Thus, it captures the main features of models that describe the gluon
propagator in the infrared,
such as a branch cut and complex poles~\cite{Binosi:2019ecz,
Falcao:2020vyr,Oliveira:2021job,Horak:2022myj}.
Note that this function is a product of a Stieltjes function, $\log(z+2)$,
with a rational meromorphic function. Therefore, the appearance of complex
Froissart doublets is not expected in its approximation by PAs~\cite{Baker1975essentials,Baker1996pade}.

We start by building genuine PAs to the Taylor expansion of $g(z)$ around $z=0$.
By applying PAs from the diagonal and near-diagonal sequences
$P_N^N(z)$, $P_{N+1}^N(z)$ and $P_N^{N+1}(z)$, a number of
features is evident.
In particular, the reproduction of the analytic aspects is hierarchical.
Indeed, the pair of complex poles and the zero closer to the origin are
already replicated by PAs of lower order ($N=2$ or 3), while the second
zero (at $z=-2.508$), which is further away from the origin,
is reproduced only by PAs with more parameters.
This is expected, since poles and zeros closer to the
origin are usually reproduced first. At the same time, the
PAs also mimic the cut by accumulating (artificial)
poles interleaved with zeros along the branch cut. This, however,
requires many parameters and is salient only when a large number of parameters
is available. Furthermore, the position of the first pole of this type tends
to approach the branch point but, again, it only leads to a good reproduction
of the branch point if sufficiently many parameters are available.

\begin{figure}[t]
\centering
\subfloat[]{\label{fig:polezerotaylor}{\includegraphics[width=0.43\textwidth]{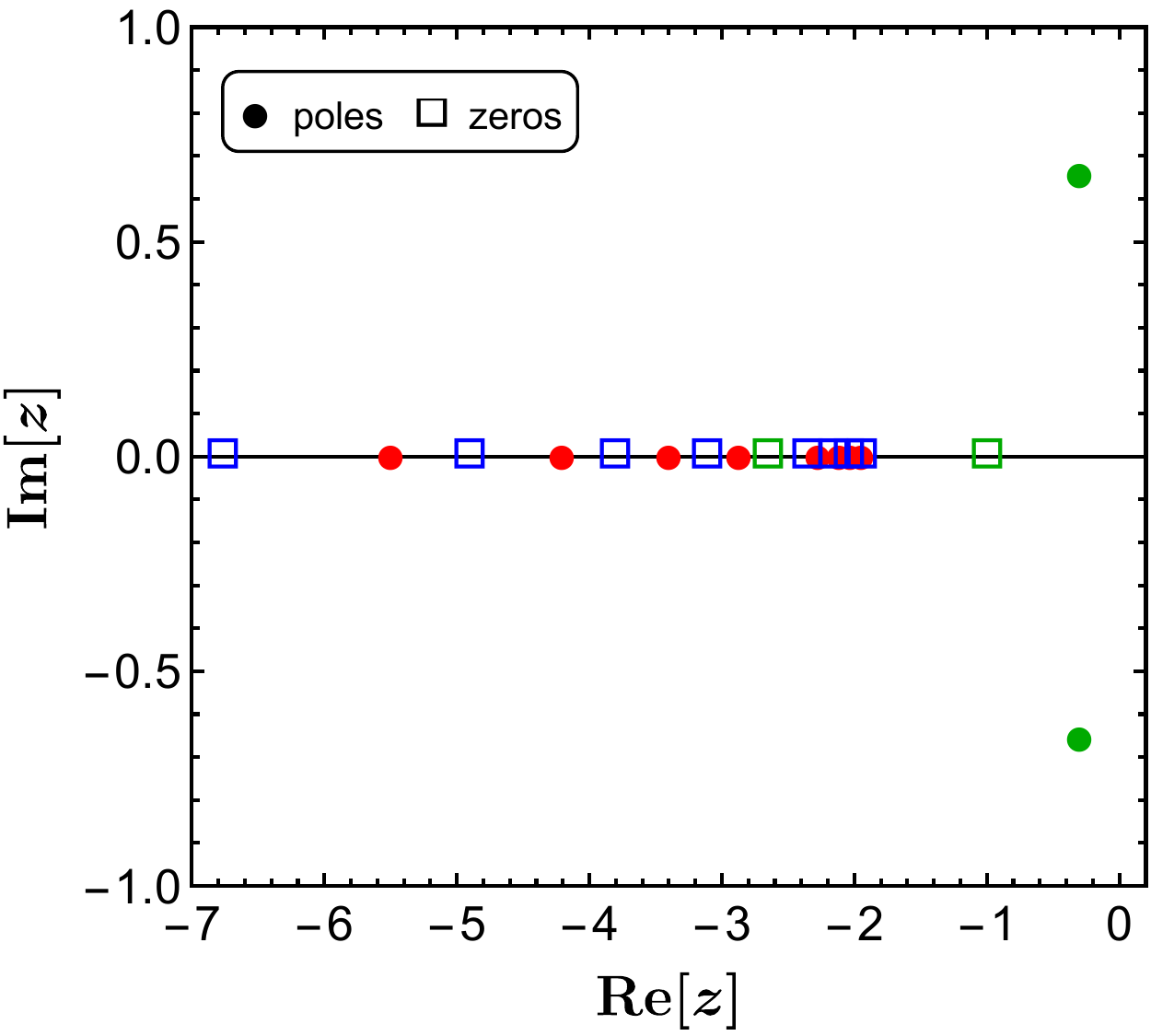}}}\hfill
\subfloat[]{\label{fig:poleconvtaylor}{\includegraphics[width=0.49\textwidth]{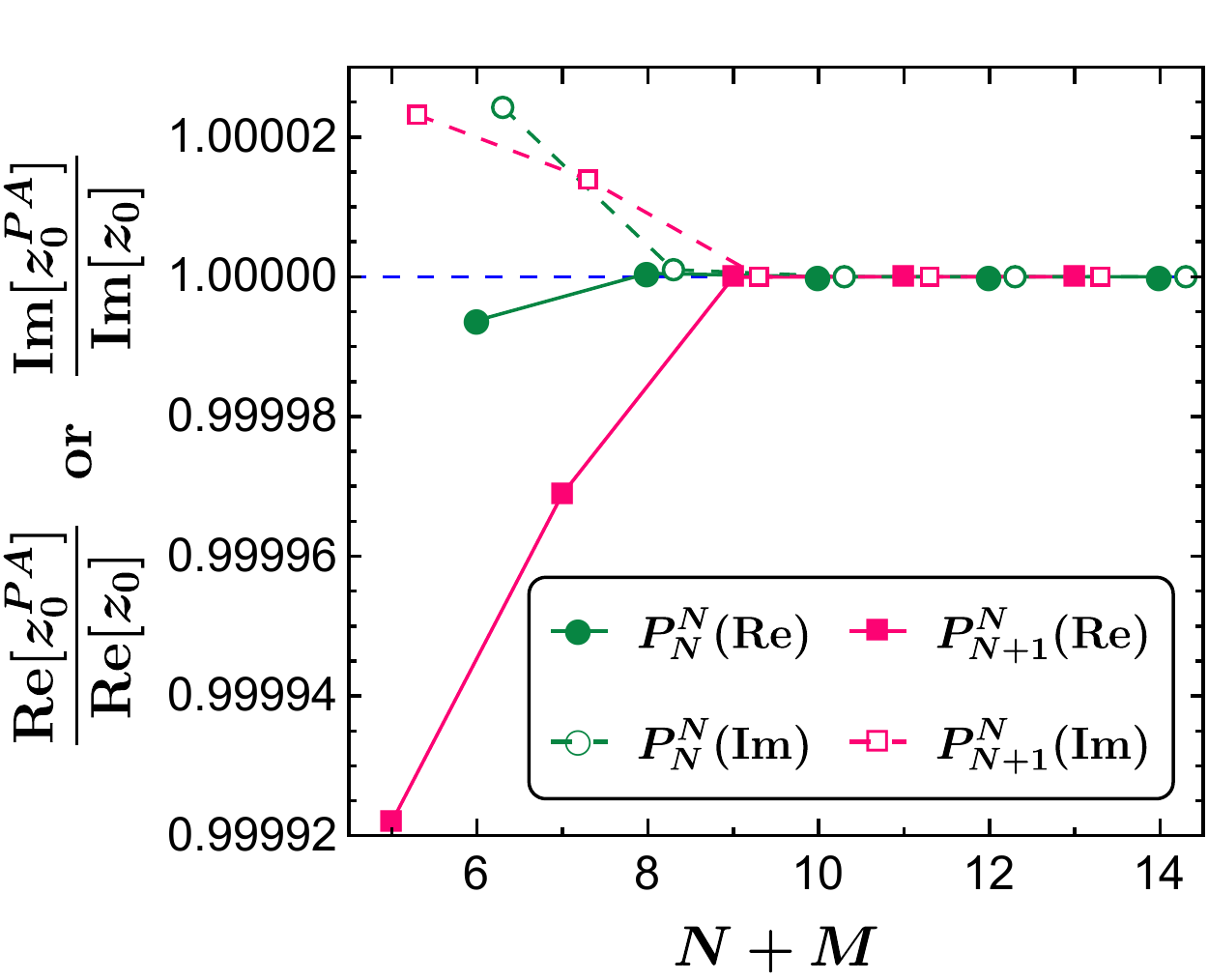}}}
\caption{(a) Poles (filled circles) and zeros (empty squares) of
$P_{15}^{14}(z)$ built to the Taylor series of $g(z)$ defined in
Eq.~(\ref{eq:logpole}). The poles and zeros that can be identified with genuine
poles of $g(z)$ are shown in green, while artifacts are shown in red (poles) and blue
(zeros). (b) Real and imaginary parts of the complex poles
$z_0^{\rm PA}$ (normalized to their true values) for
the sequences $P_N^N(z)$ and $P_{N+1}^N(z)$, as a
function of the number of parameters of the PA. To avoid inaccurate numerical
results for the higher-order PAs (see Appendix~\ref{app:Precision}), the
results in this figure were obtained working with 30 decimal places.
\label{fig:toydata}}
\end{figure}

To be concrete, in Fig.~\ref{fig:polezerotaylor}, we show the analytic structure
of a typical higher-order PA, $P^{14}_{15}(z)$. The green circles show the pair of
complex poles, which can easily be identified with the true poles of
$g(z)$ (we refer to these poles as ``physical''). The green squares show the
zeros of $g(z)$, whereas red circles and blue squares represent
artificial (or ``non-physical'') poles and zeros, respectively. The main
characteristic of the physical poles and zeros is the fact that they are
stable, i.e.\ they remain essentially fixed as $N$ increases.
The impressive convergence of the pole position obtained from the PAs to the
correct value is shown as a function of $N+M$ in Fig.~\ref{fig:poleconvtaylor},
where we display the real and imaginary parts of the physical poles of the PAs
(normalized to the exact values).
Finally, poles and zeros start to accumulate along the branch cut with branch
point $z=-2$, where the PA places a pole.
We stress that these results are expected when using
genuine PA approximants, but one should nonetheless verify if this also
happens when the PAs are used as fitting functions.
Finally, it is important to mention that when constructing Pad\'e approximants
of higher order, such as $P^{14}_{15}(z)$ of Fig.~\ref{fig:polezerotaylor},
the precision of the numerical solution becomes crucial. If insufficient
precision is used, artificial (incorrect)\footnote{Let us note that, as
said before, genuine Froissart doublets should go away when higher-order
approximants are used. Also, there may be spurious pole-zero pairs that are
due to round-off errors --- which we discuss here. The two may co-exist and it can be difficult to
distinguish them. Of course, this can be clarified by improving the numerical
precision (see Appendix~\ref{app:Precision}). As stressed in Ref.~\cite{ellipse},
the term Froissart doublets is sometimes used with different meanings:
in addition to the two instances just mentioned, spurious pole-zero pairs
may also be generated by noise in the input Taylor coefficients~\cite{ellipse,MasjuanQueralt:2010hav} and, through a similar mechanism, by uncertainties in input data, when using the
approximants as fitting functions. The latter is especially relevant for the fits presented
in this work.} pairs of complex poles with very
small residues, resembling Froissart doublets, displaying a roughly
semi-circular pattern in the complex plane, can appear in the
approximant (see Ref.~\cite{ellipse} and the discussion in Appendix~\ref{app:Precision}).

Next, in order to investigate the use of PAs as fitting functions, as well
as to test our fit procedure, we built a toy data set for $g(z)$ with 400 data
points, for every value of $z$ between 0.01 and $4$ in steps of 0.01.
We generate these data using a Gaussian distribution
around the value of $g(z)$ with 1\% error,
which is in the ballpark of the errors we have in our lattice data.
The parameters of the PA are then obtained from a $\chi^2$ fit to the toy
data.\footnote{For the $\chi^2$ minimizations reported in this paper we use {\tt Wolfram Mathematica}'s function {\tt FindMinimum} (with the method {\tt PrincipalAxis}). We checked that our results are independent of the minimization function employed. In particular, we checked that the use of the function {\tt NMinimize}, as in Ref.~\cite{Falcao:2020vyr,Oliveira:2021job}, leads to the same results.} The errors of the parameters were estimated with four different
methods: Hessian matrix,
Monte Carlo error propagation, $\Delta \chi^2$ variation (see, for example,
the Particle Data Group review on Statistics~\cite{ParticleDataGroup:2020ssz}),
and a linear error propagation (discussed in Appendix~\ref{sec:AppCorrData}).
In all cases the errors we obtain from the four methods are in good
agreement. A few observations are, however, in order.
If the number of parameters exceeds $8$ or so, the statistical uncertainties
grow dramatically and the fit parameters are not meaningful anymore.
Consequently, since the errors propagate to pole positions and
zeros, these become equally meaningless.
The bottom line is that, with our toy data set, which is similar
to our real lattice data, we must limit the number of parameters in
the PA to at most 8 or 9. (These conclusions also apply to
fits to the real data, as we discuss further below.)
An important point about the toy data is that, due to statistical fluctuations,
we can no longer guarantee that no complex Froissart doublets will
appear (see e.g.\ Ref.~\cite{ellipse}), even
though the underlying function is a product of a Stieltjes function with a
meromorphic function.

\begin{figure}[t]
\centering
\subfloat[]{\label{fig:polezerodata}{\includegraphics[width=0.46\textwidth]{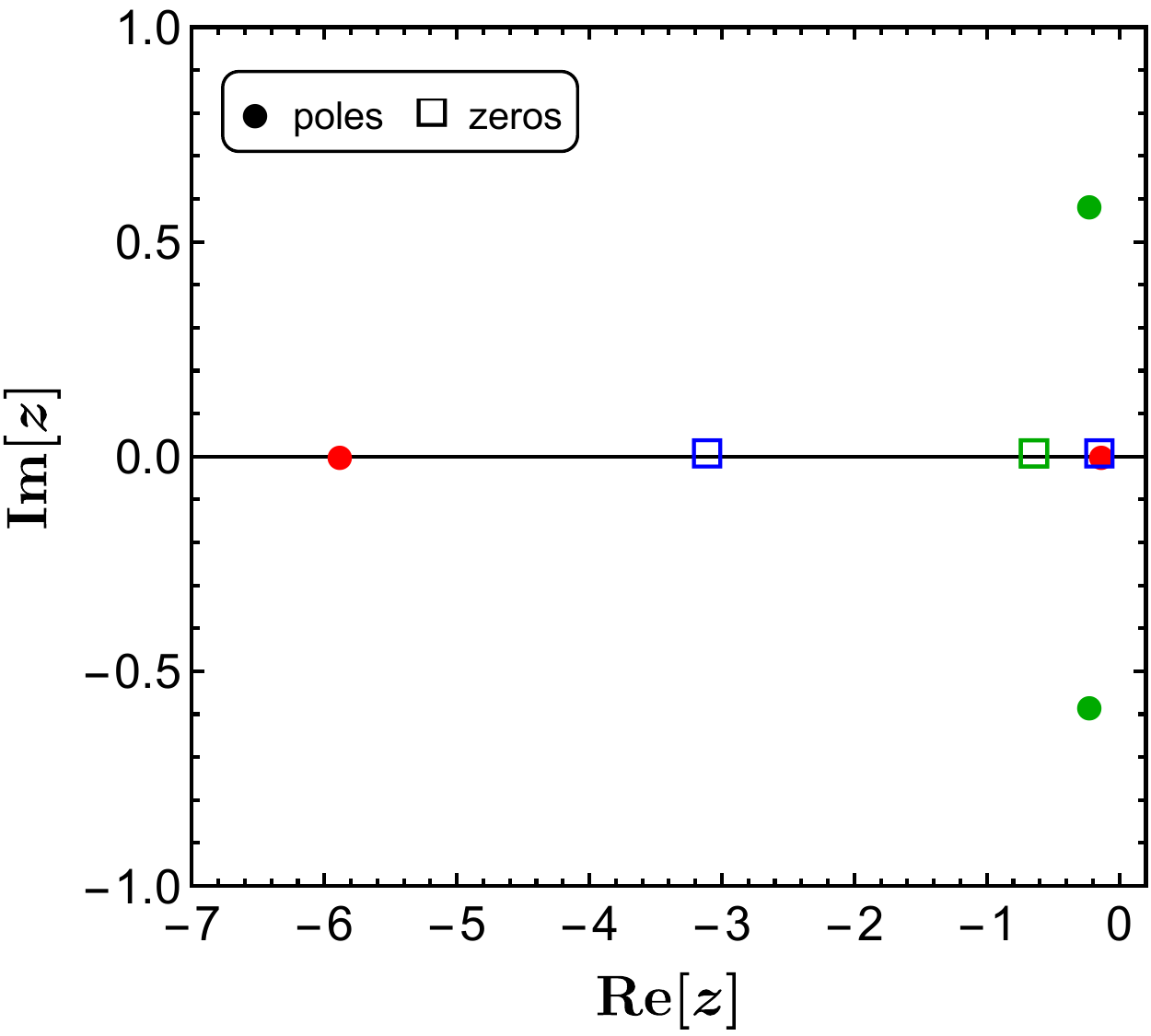}}}\hfill
\subfloat[]{\label{fig:repoleconv}{\includegraphics[width=0.48\textwidth]{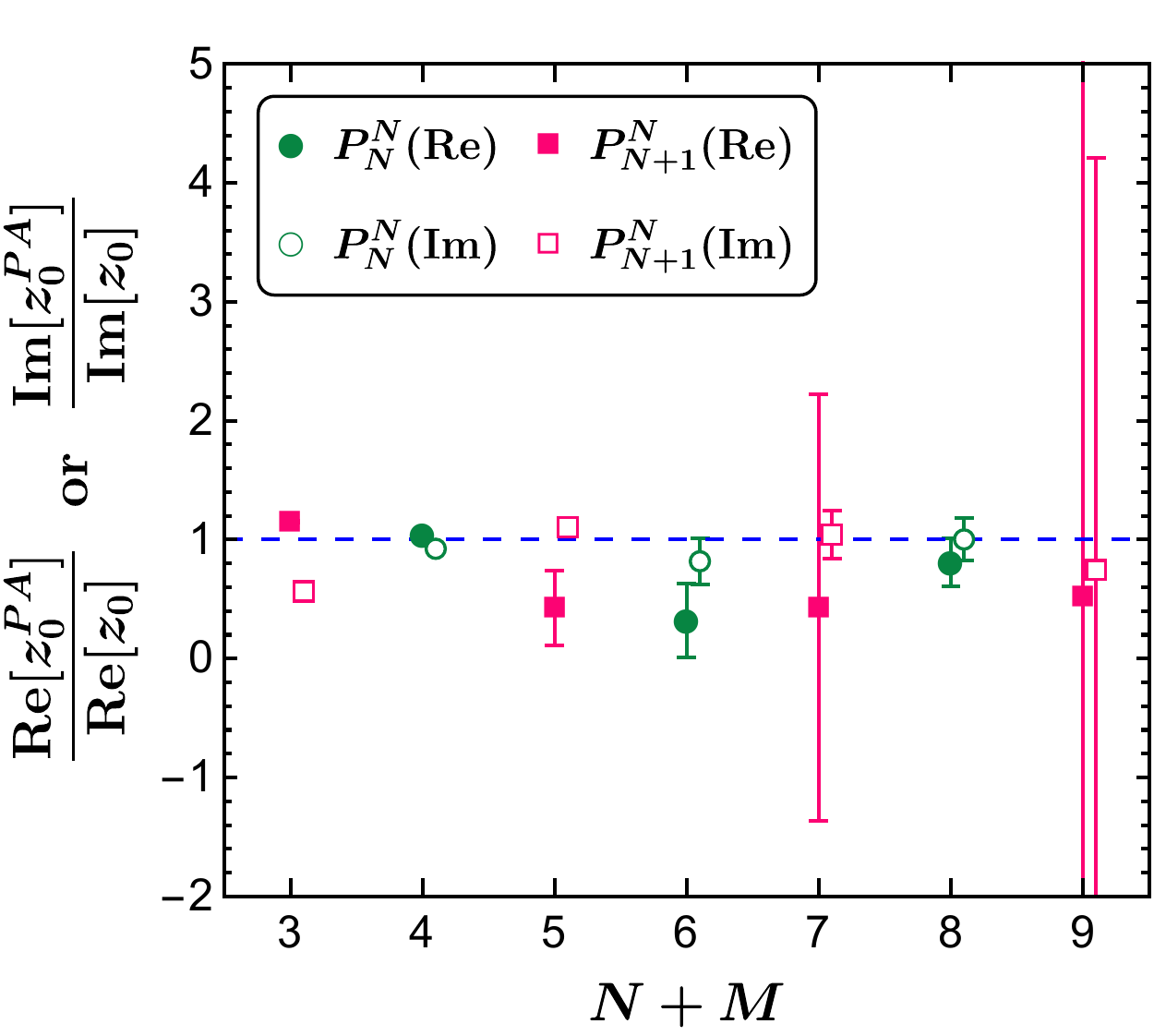}}}
\caption{(a) Poles (filled circles) and zeros (empty squares) of
$P_4^3(z)$ fitted to the toy data set generated from $g(z)$,
Eq.~(\ref{eq:logpole}). The poles and zeros that can be identified with genuine
poles of $g(z)$ are in green, while artifacts are shown in red (poles) and blue
(zeros). (b) Real and imaginary parts of the complex pole $z_0^{\rm PA}$
predicted by the PAs, normalized to their corresponding true value, for
approximants belonging to the sequences $P_N^N(z)$ and $P_{N+1}^N(z)$, as a
function of the number of parameters of the PA.}
\end{figure}

In Fig.~\ref{fig:polezerodata} we show the analytic structure of the PA
$P^3_4(z)$,
with 8 parameters, built to the considered data set.
In green, we see that the PA reproduces the poles of $g(z)$, albeit not
very precisely, and
places a zero that can be identified with the true zero at $z=-1$.
A Froissart doublet is also clearly seen, as well as a pole and a zero along
the real axis, which could be the beginning of the manifestation of the
branch cut. In Fig.~\ref{fig:repoleconv} we show the behavior of the pole
position with increasing values of $N$.
We see that the pole is relatively stable, which is very important
to identify it as a physical one.
Furthermore, it is clear that lower values of $N$ can, in fact, be
advantageous, since they lead to smaller errors in the pole position.
Finally, for more than 8 or 9 parameters, the errors on the pole position are
so large that the results are no longer meaningful.

This analysis shows that using the PAs as fitting functions
is also a reliable method to extract the main features of the analytic
structure of $g(z)$.
However, the limitations imposed by the data errors are evident: we cannot go
to very high orders in the Pad\'e sequences, and the reproduction of pole
positions and zeros has uncertainties stemming from the fit
parameters. These errors must be carefully propagated,
and grow significantly once the order of the PA is larger than $8$ or $9$.
In particular, in this situation, it becomes difficult to see the
footprints of the branch cut, which would require a larger number of
parameters.

\section{Lattice data for the Landau-gauge propagators}\label{sec:data}

The data for the (four-dimensional) $SU(2)$ Landau-gauge gluon and ghost
propagators used in the present work have been previously presented and
discussed in Refs.~\cite{Cucchieri:2007md,Cucchieri:2007rg,Cucchieri:2008fc,
Cucchieri:2009xxr,Cucchieri:2011ig,Cucchieri:2016jwg}, which contain technical
details of the simulations.
Here, we recall the main parameters and properties that are relevant for our
analysis.

We have considered symmetric lattices with various lattice sides,
allowing us to keep finite-size effects under control. In the following, we
use only the data from our largest lattice volume, $V = n^4 = 128^4$, which
can be essentially considered as infinite volume.
The lattice parameter was taken to be $\beta = 2.2$.
This yields, approximately, a lattice spacing $a$ of $0.210~\text{fm}$, obtained
by considering the input value $\sigma^{1/2} = 0.44~\text{GeV}$ for the
string tension~\cite{Bloch:2003sk}.
The resulting physical lattice volume is about $(27~\text{fm})^4$,
which is clearly much larger than a typical hadronic scale.
Correspondingly, the smallest non-zero (physical) momentum
$p_{min} = 2\,a^{-1} \sin(\pi / n )$ allowed is about $46~\text{MeV}$.

The lattice gluon propagator $D(p^2)$ in Landau gauge is evaluated using
\begin{equation}
D^{bc}_{\mu \nu}(p) \;=\; \sum_{x\mbox{,}\, y} \frac{e^{-2 \pi i \hat{p}
\cdot (x - y) / n}}{V}\, \langle A^b_{\mu}(x)\,A^c_{\nu}(y) \rangle \;=\;
\,\delta^{bc}\,\left(g_{\mu \nu}\,-\,\frac{p_{\mu}\,p_{\nu}}{p^2}\right)
D(p^2) \; .
\end{equation}
Here, $\langle \;\;\rangle$ stands for the path-integral average, $x,y$
are lattice points, $\mu,\nu$ correspond to Lorentz indices, and $b,c$
to color indices. Also, $A_{\mu}(x)$ is the lattice gluon field,
defined as $A_{\mu}(x) = [ U_{\mu}(x) - U_{\mu}^{\dagger}(x) ]/(2i)$,
where $U_{\mu}(x)$ are the usual link variables of the Wilson action.
Note that, with this definition, the gluon propagator evaluated
on the lattice corresponds to the propagator $g^2 D(p^2)$ in the
continuum~\cite{Cucchieri:1999sz}.
As for the ghost propagator $G(p^2)$, it is obtained by inverting
the Landau-gauge lattice Faddeev-Popov matrix ${\cal M}(b,x;c,y)$
--- for example, defined in Eq.\ (22) of Ref.~\cite{Cucchieri:2005yr} ---
through the relation
\begin{equation}
G^{bc}(p^2)\, = \,\sum_{x\mbox{,}\, y}
\frac{e^{- 2 \pi i \, \hat{p} \cdot (x - y) / N}}{V}\,
\langle\, {\cal M}^{- 1}(b,x;c,y) \,\rangle \;=\;
\,\delta^{bc}\, G(p^2) \; ,
\label{eq:Gdef}
\end{equation}
where $b$ and $c$ are again color indices.
For the data considered here, the matrix ${\cal M}(b,x;c,y)$ was inverted
by using a conjugate-gradient method with even/odd preconditioning
and point sources \cite{Boucaud:2005gg,Cucchieri:2006tf}.

For the ghost propagator $G(p^2)$, the data are
displayed~\cite{Cucchieri:2016jwg}
simply as a function of the (unimproved) lattice momenta
\begin{equation}
p^2 \, = \, \sum_{\mu} \, p_{\mu}^2 \; ,
\label{eq:p2}
\end{equation}
with the components $p_{\mu}$ given by $p_{\mu} = 2 \sin(\pi
\hat{p}_{\mu}/n)$, where $\hat{p}_{\mu}$ takes values $0, 1, \dots, n-1$.
Note that, since the Landau-gauge Faddeev-Popov matrix has a trivial null
eigenvalue corresponding to a constant eigenvector, one cannot evaluate
the ghost propagator at zero momentum, i.e.\ with $\hat{p}_{\mu} = 0$
for all directions $\mu$.

The data for the gluon propagator $D(p^2)$, instead, are
considered~\cite{Cucchieri:2011ig} in terms of the improved
momenta~\cite{Ma:1999kn}
\begin{equation}
p^2 \, = \, \sum_{\mu} \, p_{\mu}^2 \, + \, \frac{1}{12} \, \sum_{\mu}
\, p_{\mu}^4 \; ,
\label{eq:kim}
\end{equation}
in order to reduce effects due to breaking of rotational
symmetry, which are usually more serious at higher momenta.
This definition does not affect the value of $p^2$ in the IR limit, but
modifies its value significantly for large momenta.
In particular, the largest momentum $p_{max}$ --- obtained when
$\hat{p}_{\mu} = n/2$ for all directions $\mu$ --- is given, for the
$\beta$ value considered here, by about $3.75~\text{GeV}$ in the unimproved
case and about $4.33~\text{GeV}$ using improved momenta.

Let us stress that different lattice quantities are subject in general to
different discretization effects.
Thus, it is not surprising that gluon- and ghost-propagator data
require different definitions of the lattice momenta when one tries to connect
lattice data to the continuum analysis.
At the same time, while the improved definition in Eq.~(\ref{eq:kim}) is
expected
to reduce discretization effects for the gluon propagator $D(p^2)$, this
may not hold for functions of $D(p^2)$.
This is probably the case, for example, for the derivative of $\log D(p^2)$,
which is required in connection with the D-Log Pad\'es,
discussed in Sec.~\ref{sec:dlog}.
The same observation applies to the derivative of $\log G(p^2)$,
presented in Sec.~\ref{sec:dlog},
and which can be seen in Fig.~\ref{fig:numder1}.
However, the rather large error bars obtained for these derivatives
do not allow us to draw any conclusions regarding the breaking of
rotational symmetry (at large momenta) for these quantities.

The gluon propagator was evaluated by considering 168 independent
pure-gauge-field configurations and momenta with components $(\overline{p},
0,0,0)$, $(\overline{p},\overline{p},0,0)$, $(\overline{p},\overline{p},
\overline{p},0)$ and $(\overline{p},\overline{p},\overline{p},\overline{p})$.
Moreover, we allowed all possible permutations of the components for
momenta of the type $(\overline{p},0,0,0)$. On the contrary, we did not
consider permutations for the momenta $(\overline{p},\overline{p},
\overline{p},0)$ and in the case $(\overline{p},\overline{p},0,0)$ we
selected only permutations satisfying the constraint $p_4=0$.
At the same time, since the inversion of the Faddeev-Popov matrix
${\cal M}(b,x;c,y)$ is extremely time consuming, the data for the ghost
propagator were obtained using only a subset of 21 configurations,
among those considered for the gluon propagator.
Moreover, in this case, we considered momenta of the type $(\overline{p},0,
0,0)$,$(\overline{p},\overline{p},0,0)$, $(\overline{p},\overline{p},
\overline{p},0)$ and $(\overline{p},\overline{p},\overline{p},\overline{p})$,
with $\overline{p} > 0$ and with all possible permutations of the
momentum components $p_{\mu}$.
For both propagators --- when permutations of the components $p_{\mu}$ were
available --- an average over the different permutations was taken for each
configuration.

\begin{figure}[!t]
\centering
\subfloat[]{\label{fig:perturbative1} \raisebox{1mm}{\includegraphics[width=0.44\textwidth]{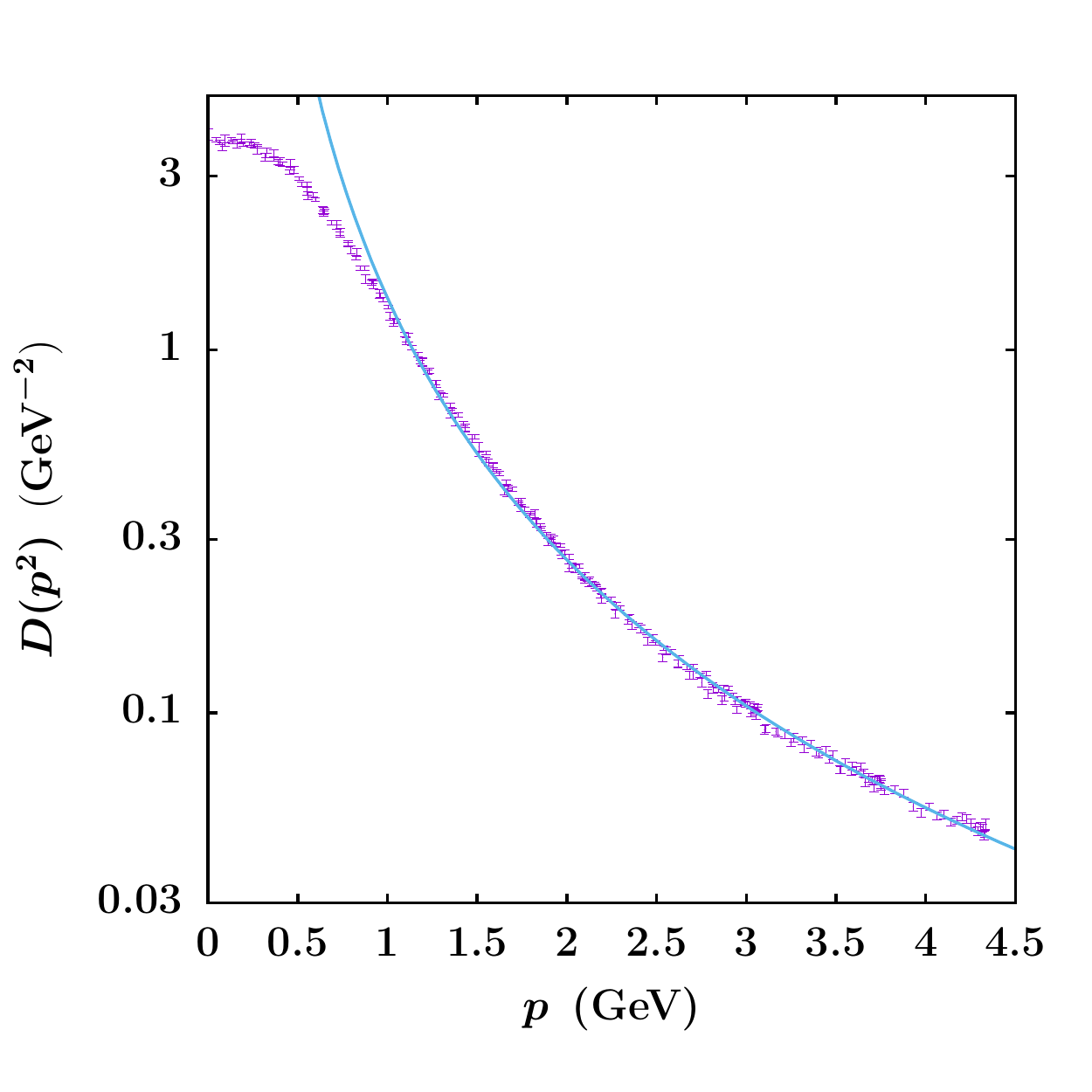}}}\hfill
\subfloat[]{\label{fig:perturbative2}{\includegraphics[width=0.45\textwidth]{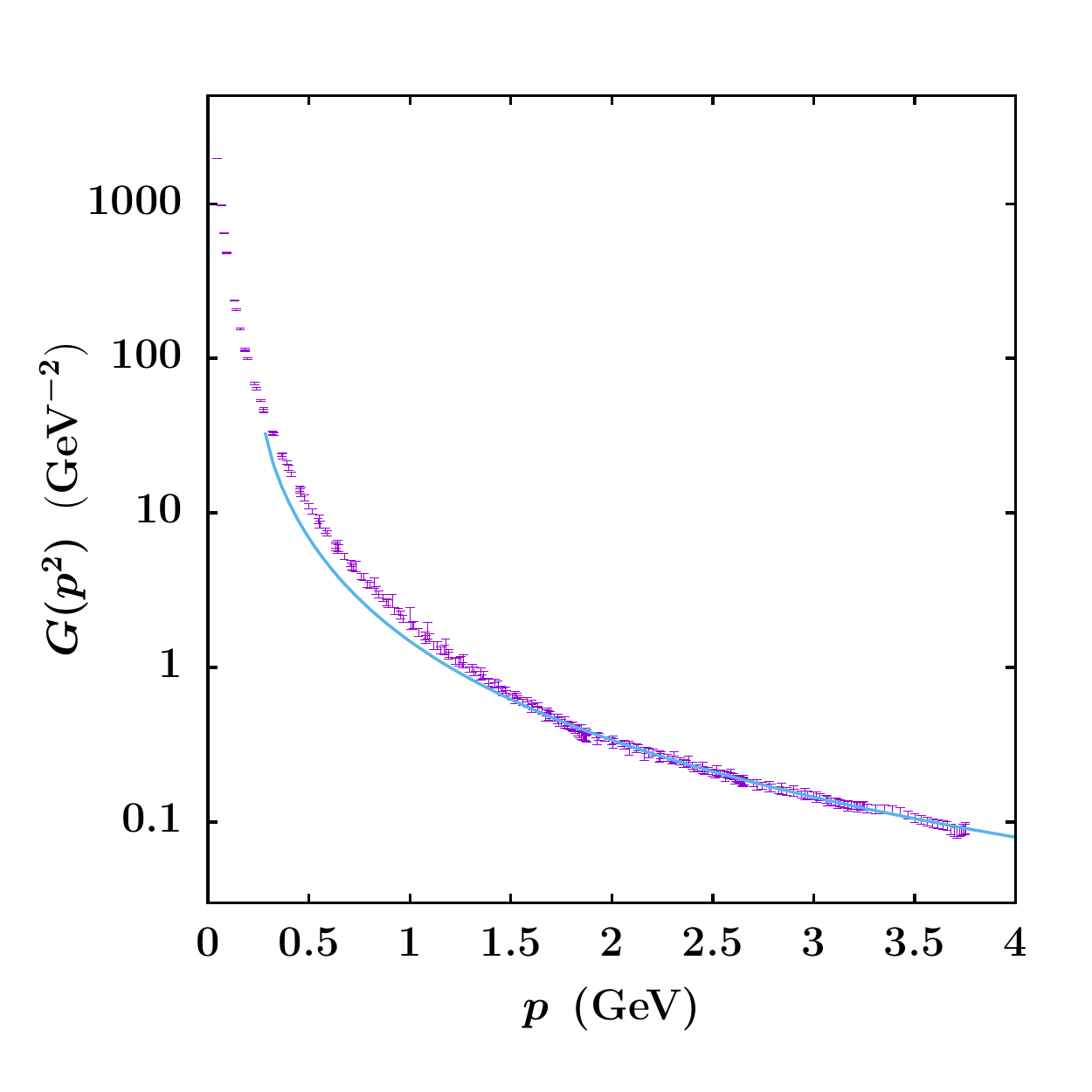}}}
\caption{\label{fig:perturbative}
Fit of the gluon (a) and ghost (b) propagators using perturbation-theory
predictions [see Eqs.~(\ref{eq:fit-gluon-UV}) and (\ref{eq:fit-ghost-UV})].
Note the logarithmic scale in the $y$ axis.
Also, recall that the variable $p$ in the $x$ axis corresponds to the
square root of $p^2$, defined in Eq.~(\ref{eq:kim}) for the gluon propagator
and in Eq.~(\ref{eq:p2}) for the ghost propagator.}
\end{figure}

Finally, we note that for momenta larger than about $1.5$--$2.0~\text{GeV}$
both propagators are essentially perturbative.
Indeed, as one can see in Fig.~\ref{fig:perturbative}, the data are
well fitted by the expression
\begin{equation}
f(p^2) \,=\, c\, \log^{-13/22}(p^2/\Lambda^2)\, / \, p^2 \; ,
\label{eq:fit-gluon-UV}
\end{equation}
for $\sqrt{p^2} \geq 2.0~\text{GeV}$ in the gluon case and by the expression
\begin{equation}
f(p^2) \,=\, c\, \log^{-9/44}(p^2/\Lambda^2)\, / \, p^2 \; ,
\label{eq:fit-ghost-UV}
\end{equation}
for $\sqrt{p^2} \geq 1.5~\text{GeV}$ in the ghost case.
[Note that we have included in the fitting functions the (one-loop)
anomalous dimensions (with $N_f =0$) for the two propagators.]
In particular, in the first case we find $c=2.31\pm 0.05$ and
$\Lambda=0.31\pm 0.03~\text{GeV}$ with $\chi^2/\mathrm{dof} = 1.73$,
while in the second one we obtain $c=1.71\pm 0.04$ and\footnote{Let
us stress that it is not immediate to compare the two values of $\Lambda$
obtained for the two fits. Indeed, these values depend strongly
on the interval of momenta used for the fit.
For example, in the ghost case, if one considers the interval $\sqrt{p^2} \geq 2.0~\text{GeV}$,
as in the gluon case, the fitted value is $\Lambda=0.262\pm 0.097~\text{GeV}$,
in reasonable agreement with the result obtained for the gluon propagator.}
$\Lambda=0.43\pm 0.07~\text{GeV}$ with
$\chi^2/\mathrm{dof} = 0.54$.
Thus, below we will focus the Pad\'e analysis mostly in the IR region.
We note that, although we are allowed
to consider at small momenta Pad\'e approximants $P^M_N(p^2)$ that do not
necessarily satisfy the leading $p^{-2}$ ultraviolet (UV) behavior of
the propagators, the sequences that have
the correct UV behavior are expected to display a faster convergence.

\section{Results}\label{sec:res}

We now turn to the use of the Pad\'e approximants of Eq.~(\ref{eq:pa}) as fitting functions to the lattice
data for gluon and ghost propagators described in Sec.~\ref{sec:data}.
The central values of the parameters are obtained through
the minimization of a $\chi^2$ function taking into account all errors and correlations (when necessary). The fit
uncertainties were calculated with several different methods (Hessian matrix, Monte Carlo error propagation, $\Delta
\chi^2$, and linear error propagation, as discussed in Sec.~\ref{sec:examples}) and we always find good
agreement between the results. For the final statistical errors we quote the values from the Hessian matrix.
The fit quality is judged by the $\chi^2$ per degree of freedom (dof) and we give the associated $p$-value in each
case. We limit our fits to approximants of relatively low orders
(up to 10 parameters), which keeps the errors
under control given the limitations imposed by the data set. Finally, when extracting pole positions, zeros, and other quantities
from the rational approximants we always propagate all the errors,
taking into account the correlations between
the fit parameters (obtained from the Hessian matrix of the fit).

The above analysis ensures a good control of the statistical errors of our
results and, as described below, it has been directly applied to the
gluon-propagator data.
However, when dealing with the ghost-propagator data, this procedure
requires some adaptations, due to the stronger correlations present in
these data.
In particular, the fit quality has to be modified, considering
only the diagonal elements of the covariance matrix, and the error
propagation is evaluated using a linear approximation,
taking into account all correlations.
This is discussed in detail in Sec.~\ref{sec:ghostprop} and in
Appendix~\ref{sec:AppCorrData}.

\subsection{Pad\'e approximants to the Landau-gauge gluon propagator}
\label{sec:gluonprop}

For the gluon propagator we
limit our fitting procedure, described above,
to approximants of relatively low orders (at most 7 parameters),
which keeps the errors under control, given the limitations imposed by the data set
(see discussion in the previous section).
As a first step, comparing
fits with and without correlations between the lattice data points,
we have checked explicitly that the correlations in the data sets are negligible
and thus can be disregarded in the fitting procedure,
in agreement with Refs.~\cite{PhysRevD.49.1585,Dudal:2019gvn}.
Then, we performed trial fits, using different numbers of data points, and compared
our results with those presented in Ref.~\cite{Cucchieri:2011ig}.
We find that, by restricting the fits to $\sqrt{p^2}<1.63~\text{GeV}$
(corresponding to 100 data points) the fit quality is very good
--- with high $p$-values --- but the obvious trade-off
is that the parameters have larger errors, due to the reduced information.
Conversely, by increasing
the number of data points, the errors become smaller but the fit quality decreases.
It turns out that by restricting the fits to $\sqrt{p^2}<2.4~\text{GeV}$ (corresponding to
160 data points) we obtain acceptable fits --- with $p$-values of the order of a
percent --- while having errors on the parameters that are small in comparison to fits
with fewer data points.
Thus, with this choice, we restrict our analysis mostly to the IR region,
which is the main focus of our study, since in this limit
the propagator is not described by the perturbative behavior.
Therefore, all the fits reported below are performed for
$\sqrt{p^2}<2.4~\text{GeV}$, with 160 data points in total.
In any case, we checked that the results do not show a strong dependence on this choice.

Before presenting our general results for the gluon propagator, we
analyze in detail, as an example, the case of a low-order approximant.
Of course, one could start with the simplest option
$P_1^1(p^2)$,\footnote{Occasionally,
in the text, as well as in tables and figures, we will simplify the
notation by omitting the $p^2$ argument in the PAs, i.e.\
denoting $P^M_N(p^2)$ by $P^M_N$.}
but fits with Pad\'es of the type $P_1^N(p^2)$ lead to very bad fits, with extremely small $p$-values.
Indeed, due to the trivial structure of their denominator, these Pad\'es are forced
to have a real pole,
which turns out to be incompatible with a precise description of the lattice data
for the gluon propagator in the IR region.
Therefore, these approximants are not included in our analysis and we
start our study with the Pad\'e $P_2^1(p^2)$, whose expression is given by
\begin{equation}
P_2^1(p^2) = \dfrac{a_0 + a_1 \, p^2}{1 + b_1 \, p^2 + b_2 \, p^4} \; . \label{eq:PAex}
\end{equation}
As already stressed above, this approximant corresponds to the function
$f_1(p^2)$ of Ref.~\cite{Cucchieri:2011ig}, which arises in the context of the
Gribov-Zwanziger scenario~\cite{GRIBOV19781,Vandersickel:2011zc,Vandersickel:2012tz}
for the description of the gluon propagator.
At the same time, note that this approximant is the first in the sequence $P_{N+1}^N(p^2)$,
which is considered here, allowing us to analyze the data in a systematic and model-independent way.
Using this Pad\'e approximant as the fitting function to perform a $\chi^2$ minimization,
we obtain $\chi^2/\mathrm{dof} = 1.28$ and a $p\mathrm{-value}$ of $0.010$, which indicates
an acceptable --- although not impressive --- fit quality.
The corresponding PA's parameters with their statistical uncertainties,
as determined by the fit, are
\begin{align}
a_0 &= (3.82 \pm 0.02) \,\,\, \mathrm{GeV}^{-2}\;, \qquad \qquad a_1 = (1.21 \pm 0.06) \,\,\, \mathrm{GeV}^{-4}\;, \nonumber \\
b_1 &= (1.18 \pm 0.02) \,\,\, \mathrm{GeV}^{-2}\;, \qquad \qquad b_2 = (1.65 \pm 0.05) \,\,\, \mathrm{GeV}^{-4}\;,
\end{align}
with the following non-trivial correlation coefficients
\begin{equation}\label{eq:covmatrixparm}
\begin{array}{c|ccc}
& a_1 & b_1 & b_2\\ \hline
a_0 & -0.426 & 0.821 & -0.436\\
a_1 & - & -0.586 & 0.995\\
b_1 & - & - & -0.632
\end{array} \;\;.
\end{equation}
We remark that some of the fit parameters are quite strongly correlated, which must be taken
into account in subsequent error propagations.

Using the fit results we can now extract the poles and the zeros of this approximant,
carefully propagating the uncertainties by taking into account the correlations between the
fit parameters.
The PA $P_2^1(p^2)$ has a pair of complex poles located at
\begin{equation}
p^2 =\left[ (-0.36 \pm 0.02) \pm i\,(0.690 \pm 0.005) \right] \,\, \mathrm{GeV}^2 \; ,
\end{equation}
and a zero on the negative real axis at
\begin{equation}
p^2 = (-3.2 \pm 0.2) \,\,\, \mathrm{GeV}^2 \; .
\end{equation}
We will see below that these complex poles and the zero are still
present in higher-order PAs, which is a strong indication of the fact that they are genuine
features of the gluon propagator in the IR and not transient artifacts.

We can also study the behavior of the propagator near $p^2\approx 0$ by
obtaining an estimate of its Taylor-series coefficients. By writing
the Taylor series of the gluon propagator around $p^2 = 0$ as
\begin{equation}
D(p^2) = c_0 + c_1 \, p^2 + c_2 \, p^4 + c_3 \, p^6 + c_4 \, p^8 + \cdots \; ,
\label{eq:seriesprop}
\end{equation}
and by expanding the Pad\'e given in Eq.~(\ref{eq:PAex})
(again taking into account the correlations in the error propagation),
we find the coefficients
\begin{align}
c_0 &= 3.82 \pm 0.02 \,\,\, \mathrm{GeV}^{-2}, & c_1 &= -3.3 \pm 0.1 \,\,\, \mathrm{GeV}^{-4}, & c_2 &= -2.4 \pm 0.4 \,\,\, \mathrm{GeV}^{-6}, \nn \\
c_3 &= 8.3 \pm 0.4 \,\,\, \mathrm{GeV}^{-8}, & c_4 &= -5.9 \pm 0.6 \,\,\, \mathrm{GeV}^{-10}, & c_5 &= -7 \pm 2 \,\,\, \mathrm{GeV}^{-12}.
\end{align}
We can see that the first few coefficients have rather small errors, which could serve as an additional
constraint to the theoretical description of the gluon propagator in the IR.

\begin{table}[!t]
\centering
\caption{Complex poles of the Pad\'e approximants used as fitting functions to the Landau-gauge gluon-propagator lattice data,
together with the $\chi^2/\mathrm{dof}$ and the p-value of the respective fits. Results from $P_2^3(p^2)$ and $P_3^3(p^2)$ do not
enter our final values (see text).}
\begin{tabular}{clcc}
\toprule
PA & complex poles \small{$(\mathrm{GeV}^2)$} & $\chi^2/\mathrm{dof}$ & p-value \\ \midrule
$P_2^1$ & $(-0.36 \pm 0.02) \pm i \, (0.690 \pm 0.005)$ & 1.28 & 0.010 \\
$P_3^1$ & $(-0.32 \pm 0.02) \pm i \, (0.65 \pm 0.03)$ & 1.27 & 0.012 \\
$P_2^2$ & $(-0.32 \pm 0.02) \pm i \, (0.65 \pm 0.02)$ & 1.28 & 0.012 \\
$P_3^2$ & $(-0.47 \pm 0.05) \pm i \, (0.66 \pm 0.03)$ & 1.19 & 0.055 \\
\rowcolor[RGB]{221,221,221}
$P_2^3$ & $(-0.27 \pm 0.01) \pm i \, (0.49 \pm 0.04)$ & 1.18 & 0.060 \\
\rowcolor[RGB]{221,221,221}
$P_3^3$ & $(-0.5 \pm 0.7) \pm i \, (0.07 \pm 3.3)$ & 1.18 & 0.068 \\ \bottomrule
\label{tab:PAspoles}
\end{tabular}
\end{table}

The results obtained using the PA $P_2^1(p^2)$ for the complex
pole, the zero, and the Taylor-series coefficients should, of course,
be corroborated by the investigation of PAs of higher orders, in the same
sequence as well as in other near-diagonal sequences.
In Table~\ref{tab:PAspoles}, we display our findings for the complex pole position
extracted from PAs belonging to the sequences $P_N^N$, $P_N^{N+1}$,
$P^{N}_{N+2}$, and $P_{N+1}^N$.
In each case we also give the value of the $\chi^2/{\rm dof}$, as well as the accompanying $p$-value.
All fits are reasonable from the point of view of the fit quality.
The parameter errors and the errors entailed in the pole position, however,
grow rapidly once the number of parameters exceeds 6.
Indeed, the PAs $P^2_3(p^2)$ and $P_3^3(p^2)$ have parameters $a_i$ and $b_i$
with errors of the order of (or larger than) 100\%, and some of the parameters are more than 99\%
correlated. Thus, these PAs have a number of parameters that
are already at the limit of what can be done with the lattice data at hand and, therefore,
we refrain from showing any results for PAs of even larger orders.
Nevertheless, somewhat surprisingly, the pole position obtained
with $P_3^2(p^2)$ has acceptable (even though larger) errors, and corroborates the
outcomes from the lower-order PAs. These not-so-large errors, in this case, are
the result of the interplay between the large parameter errors and their
strong (mostly positive) correlations, which leads to a partial cancellation
of the final error in the pole position.
On the contrary, in the case of $P_3^3(p^2)$, the large parameter errors do translate into
very large errors for the pole position, as shown in the final row of Table~\ref{tab:PAspoles}.
Because of these huge errors and correlations, we do not use $P_3^3(p^2)$ --- or $P_2^3(p^2)$, as discussed below --- for our final values.

\begin{figure}[!t]
\centering
\includegraphics[width=0.5\textwidth]{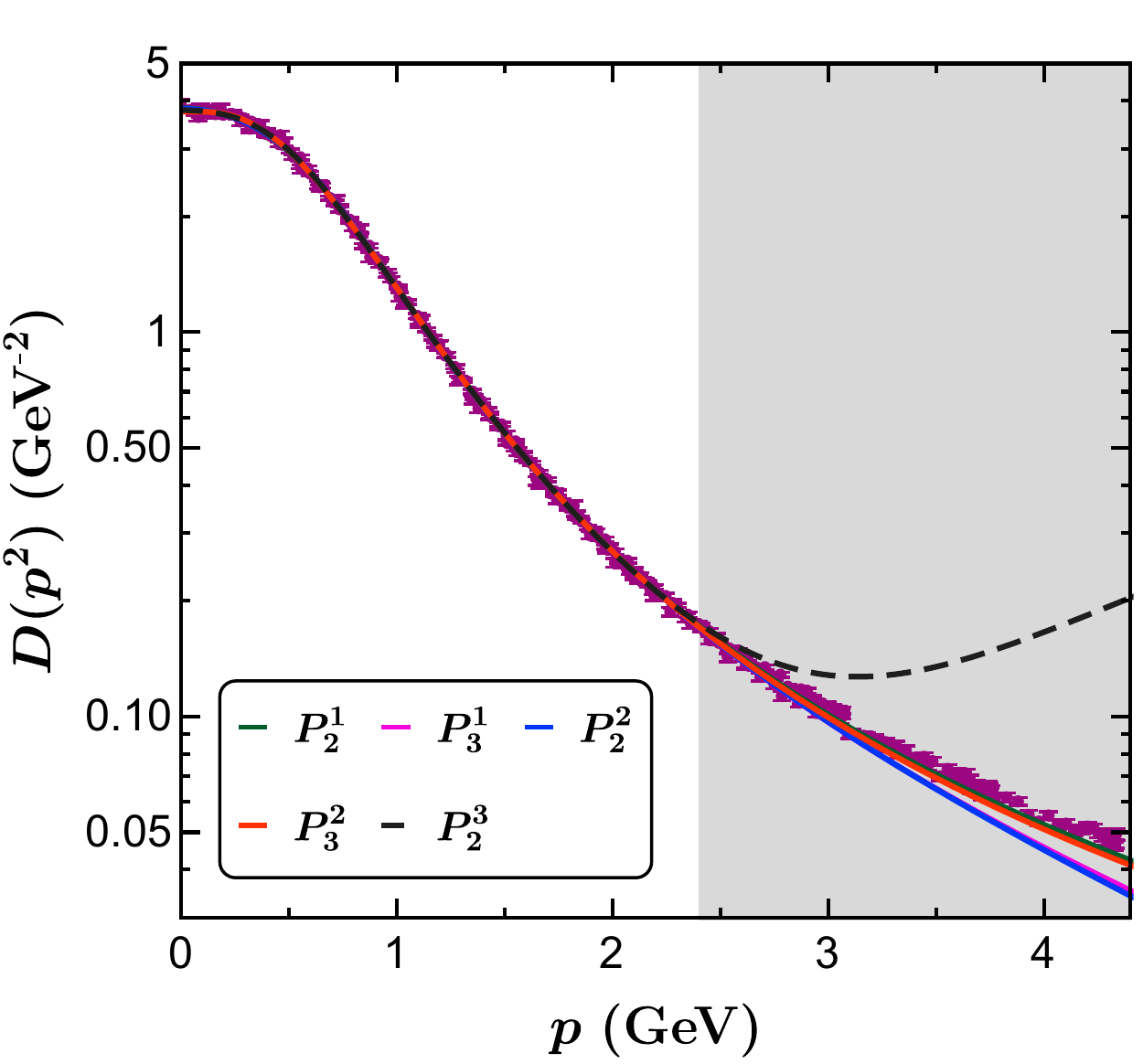}
\caption{Comparison of the Pad\'e approximants used to determine the
final results and the lattice data. The shaded region --- corresponding
to $p = \sqrt{p^2} \geq 2.4~\text{GeV}$ --- is not included in the
fits. }
\label{fig:pasdata}
\end{figure}

The results of the fits of Table~\ref{tab:PAspoles} are also
shown, compared to the lattice data, in Fig.~\ref{fig:pasdata}.
In this figure one can see the extrapolation of the PA results beyond the fit region.
In all cases, with the exception of $P^3_2(p^2)$, the PAs follow the expected behavior,
at least qualitatively.
Note that the bad UV behavior of $P^3_2(p^2)$ can be
understood, since this approximant goes as $a_4 p^2$ for large $p^2$,
with a positive --- although statistically compatible with zero
--- coefficient $a_4=0.05(20) \, {\rm GeV}^{-10}$.
Clearly, the known fall-off of the propagator in the perturbative regime
disfavors the sequences $P_N^{N+k}(p^2)$, with $k>0$, for a precise
description of the propagator in the full energy range. It is customary to exclude PAs with the wrong high-energy behavior from analyses of this type, since this will necessarily delay the convergence of the approximants~\cite{Masjuan:2007ay}.
Due to this behavior of $P^3_2(p^2)$ at higher energies, we do not include it in
our final estimate for the complex pole position, since this inclusion leads to an artificial increase of the systematic error.

\begin{figure}[!t]
\centering
\subfloat[]{\label{fig:paspoles}{\includegraphics[width=0.515\textwidth]{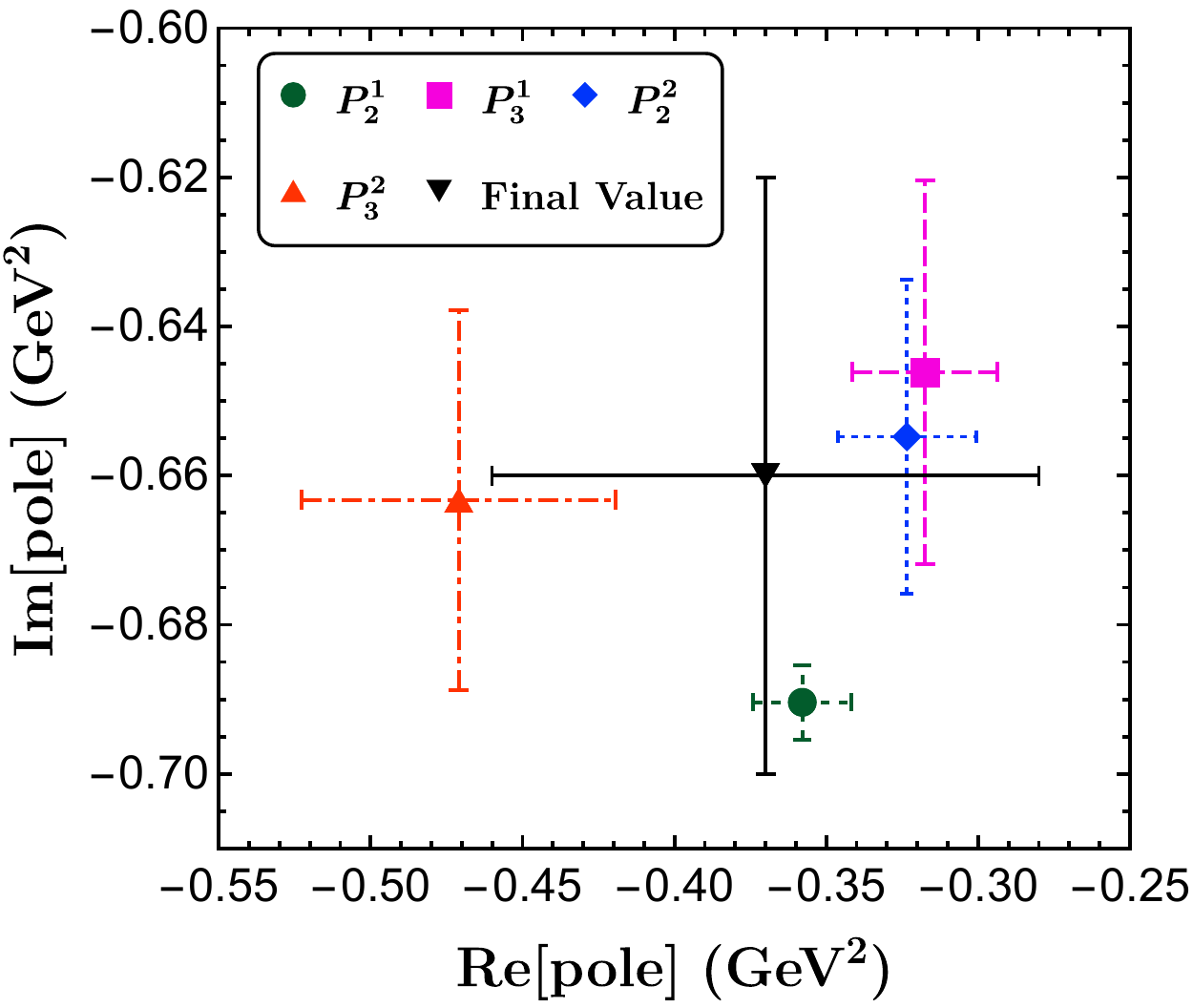}}}\hfill
\subfloat[]{\label{fig:paszeros}{\includegraphics[width=0.43\textwidth]{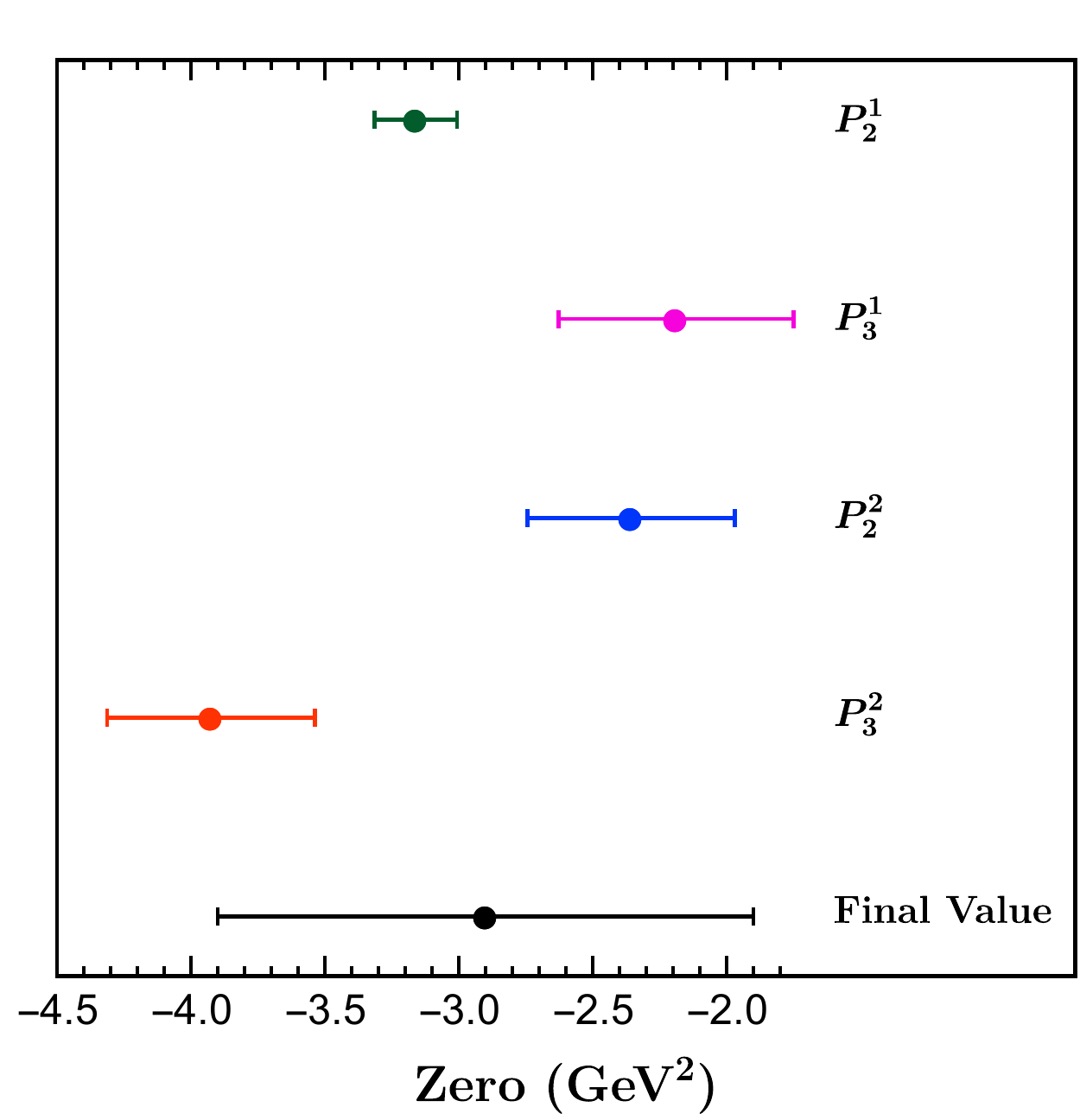}}}
\caption{Pad\'e approximants used to determine the final results.
In (a) we have the poles of each PA and in (b) the zeros. In both cases,
we show our final prediction in black. Note that the Pad\'es $P_2^1$ and $P_3^2$ belong to the same sequence $P_{N+1}^N$. Also note that in this plot we added in quadrature the two types of errors.}
\end{figure}

The results of Table~\ref{tab:PAspoles} show in all the PAs
a consistent pair of complex poles
that pass the consistency checks discussed above. We can then extract our
final value for the pole position using the results of $P^1_2$,
$P^1_3$, $P_2^2$, and $P^2_3$.
We quote as our final central value the (arithmetic) average of the results obtained
from these four PAs. Because the different results are
based on the same data set and are therefore obviously correlated,
it does not make sense to reduce the uncertainty through a weighted-average
procedure and --- to remain conservative ---
we quote, as our final statistical uncertainty, the largest one
from the individual PAs.
Finally, it is crucial to include a systematic uncertainty due to the method we employ.
We estimate this uncertainty dividing by $2$ the maximum spread between
results from two different PAs. This leads to the following value
for the position of the complex pole
\begin{align}\label{eq:polegluonfinal}
p_{\rm pole}^2 =[(-0.37 \pm 0.05_{\rm stat} &\pm 0.08_{\rm sys}) \,\, \pm \nn \\
& \pm \,\, i\,(0.66 \pm 0.03_{\rm stat} \pm 0.02_{\rm sys})] \,\, \mathrm{GeV}^2 \; ,
\end{align}
where ``stat'' and ``sys'' denote the statistical uncertainty and systematic error
from the PA method, respectively.\footnote{Note that if we include the result from $P_2^3(p^2)$ --- also shown in Table~\ref{tab:PAspoles} --- the final pole position changes very little and would read
\begin{align*}
p_{\rm pole}^2 =[(-0.35 \pm 0.05_{\rm stat} &\pm 0.10_{\rm sys}) \pm i\,(0.63 \pm 0.04_{\rm stat} \pm 0.10_{\rm sys})] \,\, \mathrm{GeV}^2 \; ,
\end{align*}
with an increased systematic error in the imaginary part. Due to the wrong UV behavior of the
sequence $P_N^{N+1}(p^2)$, this result is disfavored and our final value is that of Eq.~(\ref{eq:polegluonfinal}).}
This shows very clearly that the PAs favor the existence of a pair of complex poles,
with an imaginary part incompatible with zero. The results for the pole position in the
complex plane are summarized in Fig.~\ref{fig:paspoles}.

Another salient feature of all the PAs shown in Table~\ref{tab:PAspoles} is the presence of a zero along the real axis.
This indicates that this zero is physical and we can estimate its position using the same procedure described above
for the pair of complex poles. This leads to
\begin{equation}
p_{\rm zero}^2 = (-2.9 \pm 0.4_{\rm stat} \pm 0.9_{\rm sys}) \,\, \mathrm{GeV}^2 \; . \label{eq:pazero}
\end{equation}
The results for the zero of the PAs are shown in Fig.~\ref{fig:paszeros}.

\begin{table}[!t]
\centering
\caption{Values for the first five Taylor coefficients $c_n$ of the gluon propagator predicted by the Pad\'e approximants (with statistical errors). Shaded results have uncertainties larger than 25\% and do not enter our final values.}
\begin{tabular}{cccccc}
\toprule
& $c_0$ \small{($\mathrm{GeV}^{-2}$)} & $c_1$ \small{($\mathrm{GeV}^{-4}$)} & $c_2$ \small{($\mathrm{GeV}^{-6}$)} & $c_3$ \small{($\mathrm{GeV}^{-8}$)} & $c_4$ \small{($\mathrm{GeV}^{-10}$)} \\ \midrule
$P_2^1$ & $3.82 \pm 0.02$ & $-3.3 \pm 0.1$ & $-2.4 \pm 0.4$ & $8.3 \pm 0.4$ & $-5.9 \pm 0.6$ \\
$P_3^1$ & $3.81 \pm 0.02$ & $-3.0 \pm 0.2$ & $-3.7 \pm 0.9$ & $10 \pm 1$ & $-5.5 \pm 0.9$ \\
$P_2^2$ & $3.81 \pm 0.02$ & $-3.1 \pm 0.2$ & $-3.5 \pm 0.8$ & $10 \pm 1$ & $-5.6 \pm 0.8$ \\
$P_2^3$ & $3.78 \pm 0.02$ & $-2.1 \pm 0.4$ & $-9 \pm 2$ & \cellcolor[RGB]{221,221,221}{$23 \pm 6$} & \cellcolor[RGB]{221,221,221}{$-11 \pm 4$} \\
$P_3^2$ & $3.74 \pm 0.03$ & \cellcolor[RGB]{221,221,221}{$1 \pm 3$} & \cellcolor[RGB]{221,221,221}{$-86 \pm 109$} & \cellcolor[RGB]{221,221,221}{$1257 \pm 2653$} & \cellcolor[RGB]{221,221,221}{$-18149 \pm 54434$} \\
$P_3^3$ & $3.76 \pm 0.03$ & \cellcolor[RGB]{221,221,221}{$-0.8 \pm 1.5$} & \cellcolor[RGB]{221,221,221}{$-25 \pm 24$} & \cellcolor[RGB]{221,221,221}{$119 \pm 207$} & \cellcolor[RGB]{221,221,221}{$-397 \pm 1230$} \\ \bottomrule
\label{tab:PAscoeffs}
\end{tabular}
\end{table}

\begin{table}[t]
\centering
\caption{Final estimates for the Taylor-series coefficients of the gluon propagator. In each line, the first uncertainty is statistical while the second is systematic, obtained from the spread of values from the different PAs.}
\begin{tabular}{cl}
\toprule
$c_0$ & $(3.79 \pm 0.03_{\rm stat} \pm 0.04_{\rm sys})$ $\mathrm{GeV}^{-2}$ \\
$c_1$ & $(-2.9 \pm 0.4_{\rm stat} \pm 0.6_{\rm sys})$ $\mathrm{GeV}^{-4}$ \\
$c_2$ & $(-5 \pm 2_{\rm stat} \pm 3_{\rm sys})$ $\mathrm{GeV}^{-6}$ \\
$c_3$ & $(9.4 \pm 1.0_{\rm stat} \pm 0.9_{\rm sys})$ $\mathrm{GeV}^{-8}$ \\
$c_4$ & $(-5.7 \pm 0.9_{\rm stat} \pm 0.2_{\rm sys})$ $\mathrm{GeV}^{-10}$ \\ \bottomrule
\label{tab:finalcoeff}
\end{tabular}
\end{table}

\begin{figure}[t]
\centering
\includegraphics[width=0.5\textwidth]{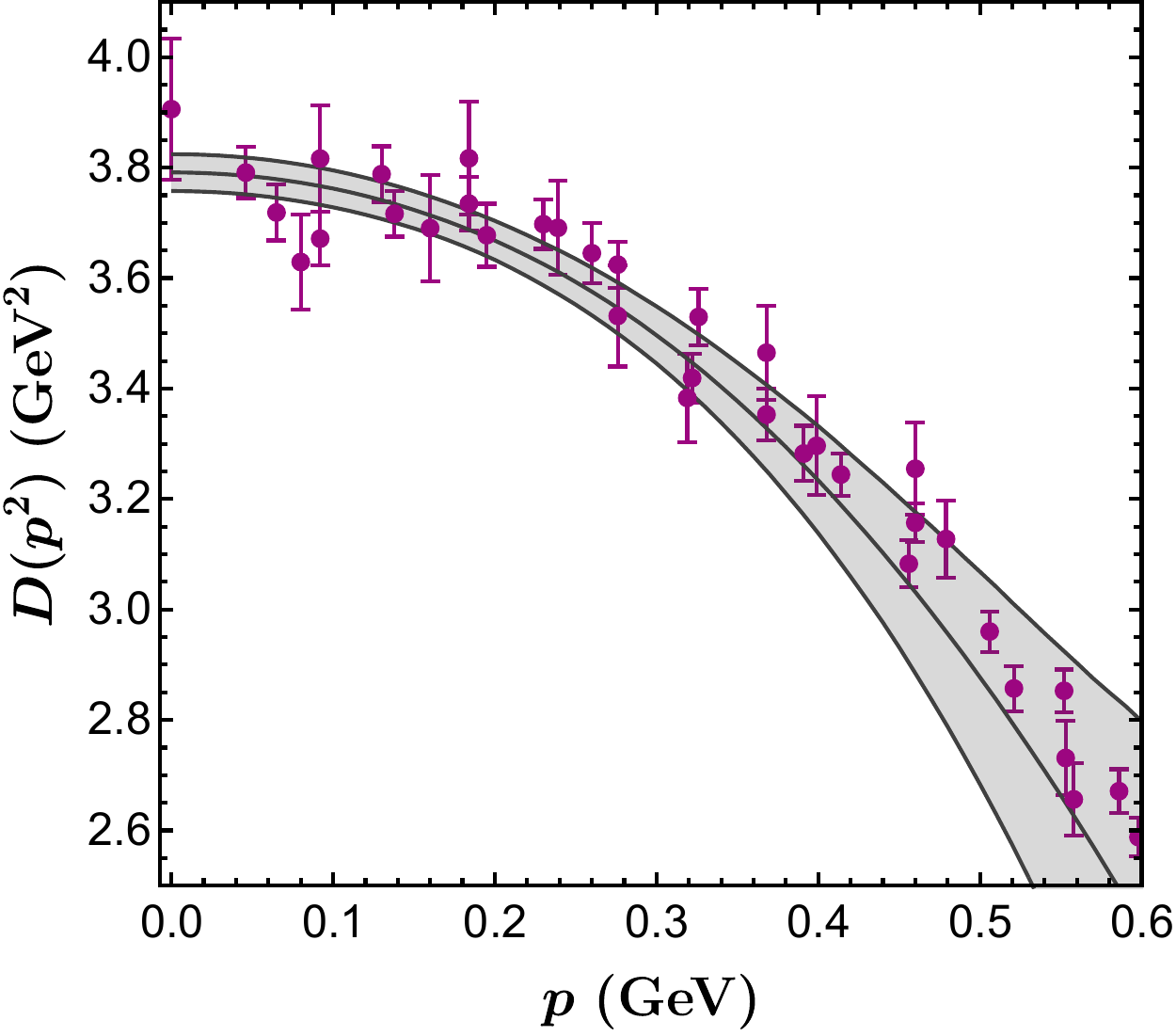}
\caption{Landau-gauge gluon-propagator Taylor series (gray) up to fifth order, from the results of Table~\ref{tab:finalcoeff}. The uncertainty band is the 68\% confidence level, assuming a uniform distribution for the
Taylor-series coefficients. The lattice data from Ref.~\cite{Cucchieri:2007rg,Cucchieri:2007md,Cucchieri:2008fc} are shown in purple.}
\label{fig:paseries}
\end{figure}

We now turn to the determination of the Taylor coefficients for the
expansion of the Landau-gauge gluon propagator around $p^2=0$.
The results for the first five Taylor coefficients obtained with the
PAs are shown in Table~\ref{tab:PAscoeffs}. In all cases,
the constant $c_0$ is very well determined. The statistical uncertainties grow with the order of the coefficients $c_n$,
as well as with the order of the PAs.
We include in our final determination of the coefficients $c_n$ the results with relative statistical uncertainty smaller
than 25\%, as shown in Table~\ref{tab:PAscoeffs}.
We follow again the procedure outlined above, i.e.\ we obtain the final Taylor coefficients from
the average of the central values.
Also, the final statistical uncertainty is the largest among the individual
PAs that contribute to the final value, and the systematic uncertainty is half of the maximum spread between
the individual determinations. The final results are shown in Table~\ref{tab:finalcoeff}. The final values for the
$c_n$ coefficients, up to order four, are then used to plot the propagator obtained from the Taylor expansion,
Eq.~(\ref{eq:seriesprop}), compared with the data points. This result is shown in Fig.~\ref{fig:paseries} and represents
a model-independent constraint on the behavior of the lattice gluon propagator for
$p=\sqrt{p^2}< 0.6~\text{GeV}$.
The error band of Fig.~\ref{fig:paseries} is obtained from a Monte Carlo error propagation by randomly generating 5000
values for each of the Taylor coefficients, assuming that they follow a uniform distribution, since the errors are
dominated by the systematic component.
The central value is then the median of the 5000 values obtained for the
Taylor series and the uncertainty band
represents the 68\% confidence level interval.

\subsection{Pad\'e approximants to the Landau-gauge ghost propagator}
\label{sec:ghostprop}

We now turn to the analysis of the Landau-gauge ghost propagator.
The method is the same as in the previous section,
which allows for a streamlined discussion of the results.
The main difference with respect to the gluon-propagator case
concerns the correlations in the data sets
from different configurations.
Indeed, for
the ghost-propagator data, these correlations are substantial.\footnote{We
note, however, that this is not the case in Ref.~\cite{Dudal:2019gvn}.}
As a matter of fact, the correlation matrix has several
non-diagonal entries with values as large as 0.75. Fits to strongly correlated data are known to be
problematic, since they can lead to biased~\cite{DAgostini:1993arp},
or simply unreliable, results~\cite{Boito:2011qt}. In particular,
in strongly correlated data, the correlation
(or covariance) matrix has very small eigenvalues.
As a consequence, its numerical inversion is problematic and
its inverse contains
very large numbers, which contribute with
large and quasi-random fluctuations in the $\chi^2$ value,
spoiling the fit results. In this situation, one often resorts to fits where the off-diagonal
correlations do not enter the fit quality,
simply by considering a diagonal covariance matrix.
We refer to fits of this type as ``diagonal fits''~\cite{Boito:2011qt}
and, in this case, we denote the fit quality by ${Q}^2$,
instead of the usual $\chi^2$.
Thus, for the analysis of the ghost propagator,
we will perform diagonal fits. We will, however,
include all correlations in the error propagation, following the procedure of
Ref.~\cite{Boito:2011qt}, which is described in Appendix~\ref{sec:AppCorrData}.
This method allows for unbiased fits with reliable errors.
On the other hand, since we use a diagonal covariance matrix,
we do not have a strict statistical interpretation for the measure of the fit
quality given by $Q^2/{\rm dof}$. Therefore, we cannot
judge the fit quality in absolute terms, but only by comparison with other
similar fits. For this reason, we do not quote $p$-values
in association with $Q^2$-type fits.

\begin{table}[t]
\centering
\caption{Location of the pole, on the negative real axis of $p^2$,
from the partial Pad\'e approximants --- used as fitting functions
to the Landau-gauge ghost-propagator lattice data --- together
with the corresponding $Q^2/\mathrm{dof}$
value. Results from $\mathbb{P}_4^4(p^2)$
and $\mathbb{P}_5^4(p^2)$ do not enter our final values,
due to their large errors.}
\begin{tabular}{ccc}
\toprule
PPA & pole \small{$(\mathrm{GeV}^2)$} & $Q^2/\mathrm{dof}$ \\ \midrule
$\mathbb{P}_1^1$ & $-0.30 \pm 0.05$ & 0.65 \\
$\mathbb{P}_2^1$ & $-0.33 \pm 0.05$ & 0.60 \\
$\mathbb{P}_1^2$ & $-0.33 \pm 0.05$ & 0.60 \\
$\mathbb{P}_2^2$ & $-0.32 \pm 0.05$ & 0.59 \\
$\mathbb{P}_3^1$ & $-0.27 \pm 0.04$ & 0.54 \\
$\mathbb{P}_3^3$ & $-0.29 \pm 0.04$ & 0.32 \\
$\mathbb{P}_4^3$ & $-0.28 \pm 0.03$ & 0.32 \\
$\mathbb{P}_5^3$ & $-0.24 \pm 0.05$ & 0.28 \\
\rowcolor[RGB]{221,221,221}
$\mathbb{P}_4^4$ & $-0.1 \pm 0.1$ & 0.27 \\
\rowcolor[RGB]{221,221,221}
$\mathbb{P}_5^4$ & $-0.16 \pm 0.08$ & 0.25 \\ \bottomrule
\label{tab:PAsghostpoles}
\end{tabular}
\end{table}

We start again our analysis with the Pad\'e approximant $P_2^1(p^2)$, since the PAs
with fewer parameters led to very bad fits. All fit parameters in this
case turn out to be of the order of $10^{10}$, which can
be seen as a manifestation of the existence of a pole very
close to the origin, namely at $p^2\approx
-1.2\times 10^{-10}~\text{GeV}^2$.
The appearance of a pole at $p^2\approx 0$ and of very large fit parameters
occurs for higher-order PAs as well.
This is a clear indication that this pole at the origin is a physical
one and thus must be explicitly enforced in the structure of the PAs,
in order to obtain fit parameters of a natural size.
This can be done using the so-called partial Padé approximants
(PPAs)~\cite{diaz2005convergence} that we define as
\begin{equation}
\mathbb{P}^M_{N,k}(p^2) = \frac{Q_M(p^2)}{R_N(p^2) \, T_k(p^2)} \; ,
\end{equation}
where the polynomials $Q_M(p^2)$ and $R_N(p^2)$ are the same as before and
in $T_k(p^2)$ we impose the existence of $k$ poles. In particular, here,
we use simply $k=1$ with
\begin{equation}
T_1(p^2)=p^2 \; ,
\end{equation}
which imposes the pole at $p^2=0$. In order to
simplify the notation, we denote these PPAs
$\mathbb{P}^M_{N,1}(p^2)$ just as $\mathbb{P}^M_{N}(p^2)$.
Let us note that an alternate procedure, which leads to the same
results, is to fit the data for $p^2\, G(p^2)$, effectively removing the
singularity at the origin.

As a preliminary step, we have again investigated fits using
different fit windows. The results we report below
are for fits restricted to the interval $\sqrt{p^2} \leq
3.12~\text{GeV}$, which contains 220 data points and enters the region that
is well described by the perturbative prediction, as shown in Fig.~\ref{fig:perturbative2}.
(We stress, however, that our main results are largely independent
of the chosen fit window.) The primary reason for our final decision is
that this choice avoids, in some of the PPAs, the appearance of
Froissart doublets located in the region where lattice data are
available. We recall that these doublets are expected to appear, in a transient form, in some of the approximants
and that they are harmless, as long as we use
the PPAs away from these singularities. However, they can spoil the extrapolation of the fit results beyond the fit window.

\begin{figure}[t]
\centering
\subfloat[]{\label{fig:paspolesghost}{\includegraphics[width=0.49\textwidth]{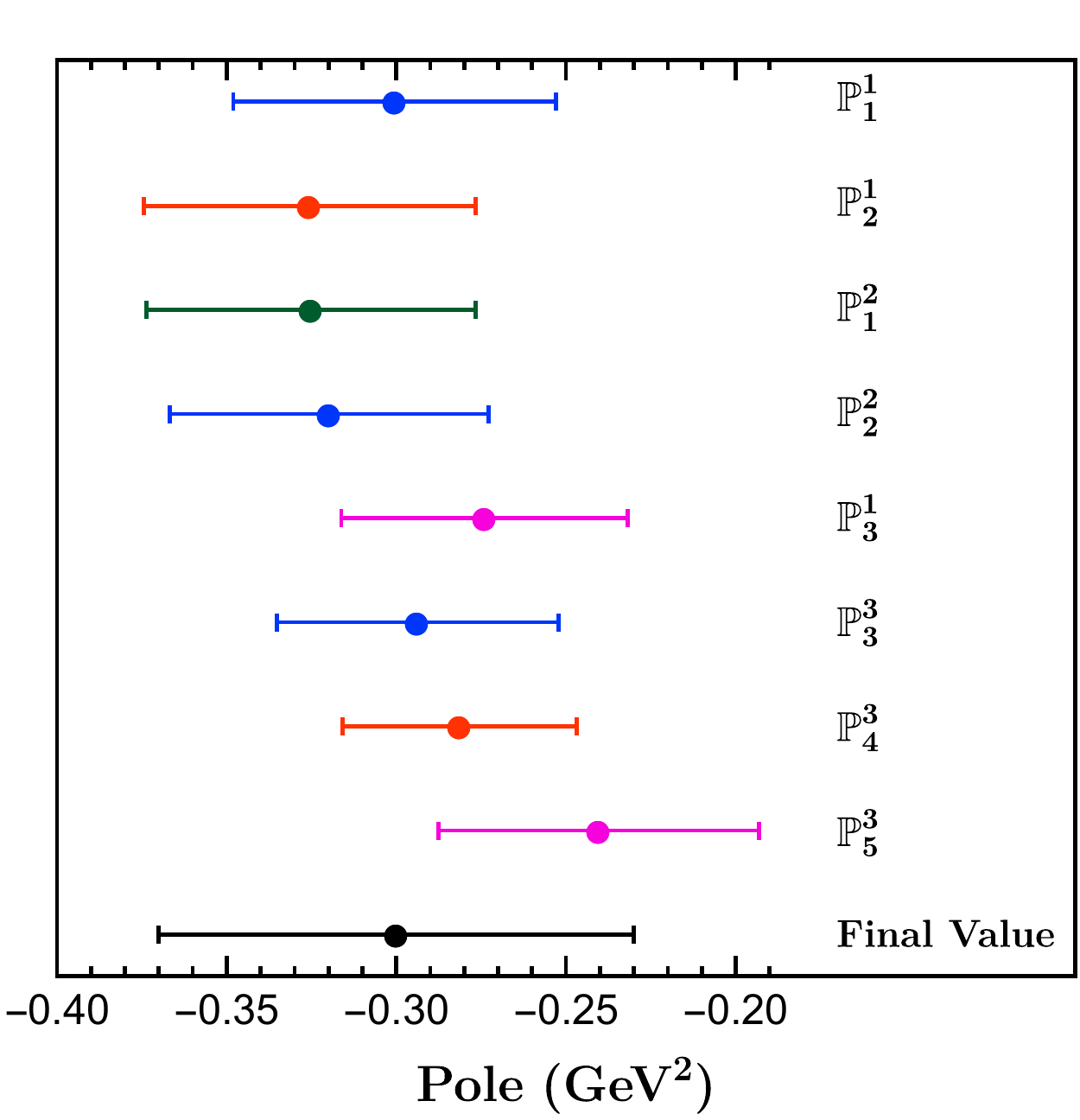}}}\hfill
\subfloat[]{\label{fig:paszerosghost}{\includegraphics[width=0.49\textwidth]{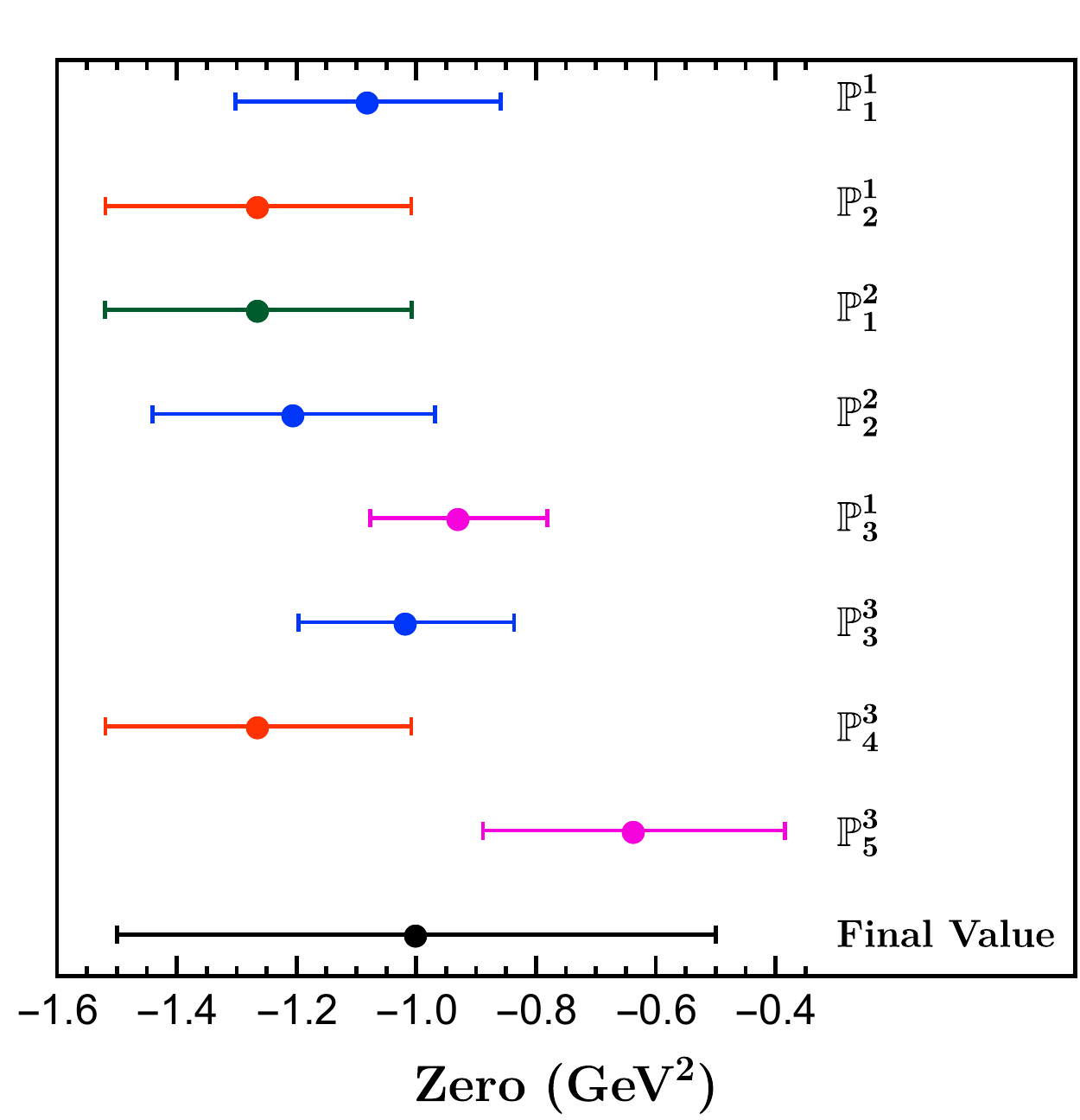}}}
\caption{Location of the (a) pole and (b) zero from the partial Pad\'e approximants, fitted to the Landau-gauge ghost-propagator lattice data. Results from approximants belonging to the same sequence are shown in the same color [blue for $\mathbb{P}^N_N(p^2)$, orange for $\mathbb{P}^N_{N+1}(p^2)$ and pink for $\mathbb{P}^N_{N+2}(p^2)$]. Our final values are shown in black at the bottom of each plot. As before, in this plot we add in quadrature the statistical and systematic errors. The pole followed by a zero points towards the presence of a branch cut along the negative real axis of $p^2$. (See discussion in the text.)}
\label{fig:pole-zero-ghost}
\end{figure}

We employed partial Pad\'es, belonging to the sequences $\mathbb{P}_N^N$,
$\mathbb{P}_N^{N+1}$, $\mathbb{P}_{N+1}^N$, and $\mathbb{P}_{N+2}^N$, to fit the ghost-propagator data.
Here, the approximants that play a special role are those belonging to the sequence
$\mathbb{P}^N_N(p^2)$, since they have a built-in $1/p^2$ behavior at
large $p^2$, which should lead to a faster convergence.
We note that a salient feature of all approximants is the appearance
of a pole located on the negative real axis of $p^2$. These results are shown
in Table~\ref{tab:PAsghostpoles}, together with the associated values of $Q^2/\mathrm{dof}$
from each fit.
Clearly, the position of this pole on the negative real axis is
remarkably stable, which strongly suggests that it has a physical nature.
At the same time, approximants with more than 8 parameters,
such as $\mathbb{P}_4^4(p^2)$ and $\mathbb{P}_5^4(p^2)$
in Table~\ref{tab:PAsghostpoles}, have strongly correlated fit parameters, of the order
of 99\%, and fit-parameter errors larger than 100\%.
Since such large uncertainties lead to predictions with
substantial errors, these PAs will not
enter our final estimates. We stress that a similar behavior
has already been observed for the PAs with more than 6 parameters in
the analysis of the gluon-propagator data.

\begin{figure}[t]
\centering
\includegraphics[width=0.5\textwidth]{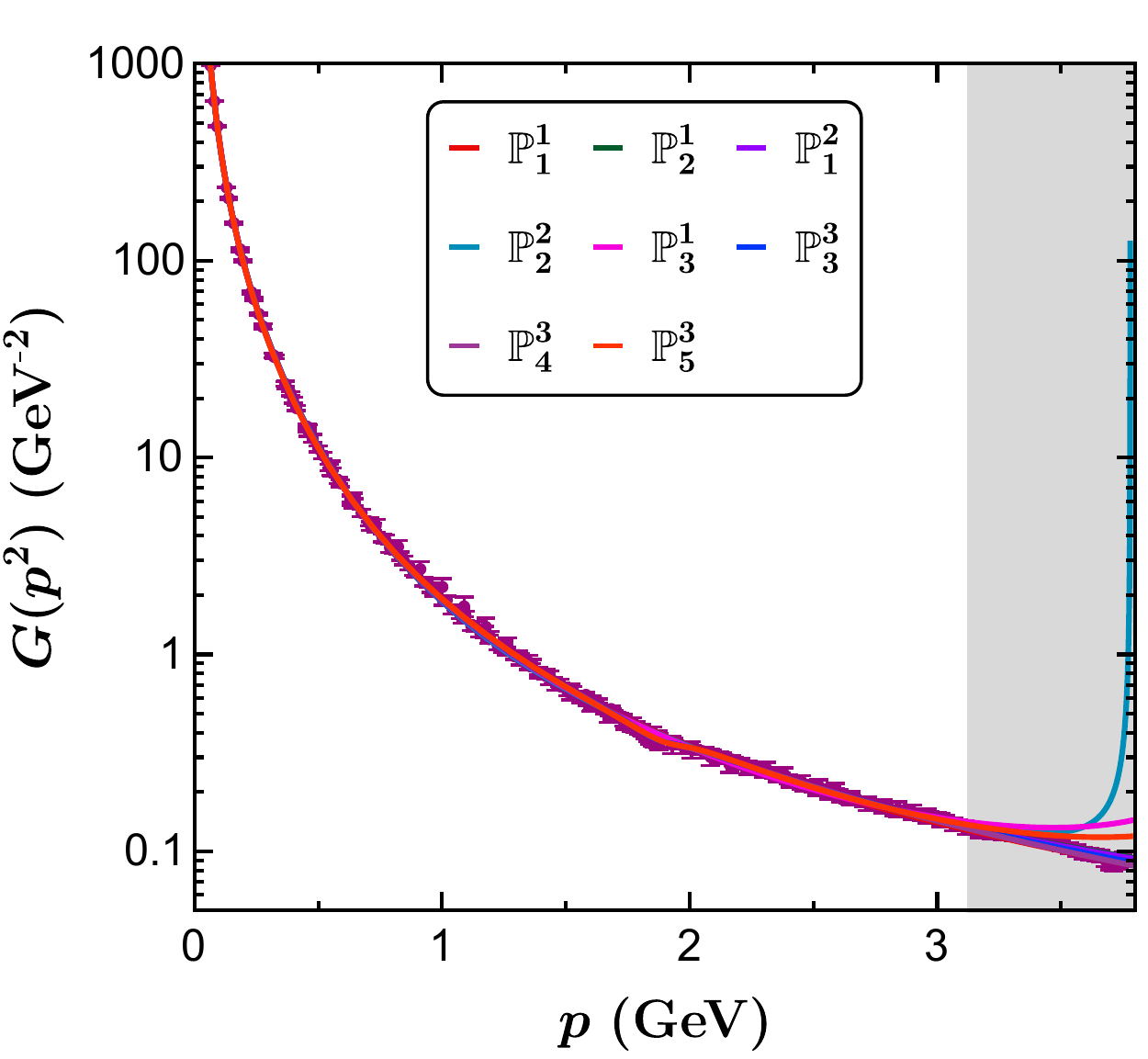}
\caption{Plot of the partial Pad\'e approximants,
used to determine our final results, and of
the lattice data for the Landau-gauge
ghost propagator. The shaded region -- corresponding
to $p = \sqrt{p^2} > 3.12~\text{GeV}$ --- is not included in the fits.}
\label{fig:pasdataghost}
\end{figure}

From the results shown in Table~\ref{tab:PAsghostpoles}, we can obtain a final
estimate for the location of the
pole predicted by the PPAs, by employing the same procedure used for the
gluon-propagator analysis: the central value is the mean of all the
results obtained from the PPAs,
the statistical uncertainty is the largest error from a single Pad\'e, and the
systematic uncertainty is half the maximum difference of two PPA results.
Thus, our final estimate for the pole position on the real axis is
\begin{equation}\label{eq:pologhostfinal}
p^2_{\mathrm{pole}} = (-0.30 \pm 0.05_{\mathrm{stat}} \pm 0.05_{\mathrm{sys}}) \,\, \mathrm{GeV}^2 \; .
\end{equation}
The results for the pole position of individual PPAs, as well as our final result,
are summarized in Fig.~\ref{fig:paspolesghost}.
At the same time,
the fitted approximants also suggest the existence of
a zero on the negative real axis of $p^2$, which, by applying the same method as
before, is predicted to be at
\begin{equation}
p^2_{\mathrm{zero}} = (-1.0 \pm 0.3_{\mathrm{stat}} \pm 0.4_{\mathrm{sys}}) \,\, \mathrm{GeV}^2 \; .
\end{equation}
The location of the zeros from the different PPAs are shown in Fig.~\ref{fig:paszerosghost}
(together with our final estimate).
It is important to remark that the existence of a pole followed by a zero
along the negative real axis may be, in fact, the manifestation of a branch
cut --- a possibility that we discuss further below.

The results from the PPAs considered in Table~\ref{tab:PAsghostpoles}
and in Fig.~\ref{fig:pole-zero-ghost}
are exhibited in Fig.~\ref{fig:pasdataghost},
compared to the lattice data for $G(p^2)$; the shaded area indicates the region not
included in the fit window. We note that the approximant $\mathbb{P}_2^2(p^2)$
has a Froissart doublet, i.e.\ a pole partially cancelled by a nearby zero, on the positive real axis,
but located outside the fit region. As said above, this is, typically, a transient
artifact of the rational approximants.
Indeed, the doublet disappears when the order of the PPA is increased and it is already
not present in the approximant $\mathbb{P}^3_3(p^2)$.

\begin{table}[t]
\centering
\caption{Values for the first five Taylor coefficients $r_n$ of the ghost propagator predicted by the partial Pad\'e approximants (with statistical errors). Shaded results have uncertainties larger than 45\% and do not enter our final values.}
\begin{tabular}{cccccc}
\toprule
& $r_0$ & $r_1$ \small{($\mathrm{GeV}^{-2}$)} & $r_2$ \small{($\mathrm{GeV}^{-4}$)} & $r_3$ \small{($\mathrm{GeV}^{-6}$)} & $r_4$ \small{($\mathrm{GeV}^{-8}$)} \\ \midrule
$\mathbb{P}_1^1$ & $4.17 \pm 0.01$ & $-10 \pm 1$ & $33 \pm 10$ & \cellcolor[RGB]{221,221,221}{$-111 \pm 51$} & \cellcolor[RGB]{221,221,221}{$369 \pm 228$} \\
$\mathbb{P}_2^1$ & $4.17 \pm 0.01$ & $-9 \pm 1$ & $29 \pm 8$ & $-89 \pm 39$ & \cellcolor[RGB]{221,221,221}{$275 \pm 161$} \\
$\mathbb{P}_1^2$ & $4.17 \pm 0.01$ & $-9 \pm 1$ & $29 \pm 8$ & $-90 \pm 39$ & \cellcolor[RGB]{221,221,221}{$276 \pm 161$} \\
$\mathbb{P}_2^2$ & $4.17 \pm 0.01$ & $-10 \pm 1$ & $30 \pm 8$ & $-94 \pm 40$ & \cellcolor[RGB]{221,221,221}{$293 \pm 168$} \\
$\mathbb{P}_3^1$ & $4.17 \pm 0.01$ & $-10 \pm 1$ & $38 \pm 10$ & $-138 \pm 58$ & \cellcolor[RGB]{221,221,221}{$503 \pm 288$} \\
$\mathbb{P}_3^3$ & $4.17 \pm 0.01$ & $-10 \pm 1$ & $34 \pm 9$ & $-117 \pm 48$ & \cellcolor[RGB]{221,221,221}{$398 \pm 221$} \\
$\mathbb{P}_4^3$ & $4.17 \pm 0.01$ & $-10 \pm 1$ & $37 \pm 9$ & $-131 \pm 47$ & \cellcolor[RGB]{221,221,221}{$466 \pm 225$} \\ \bottomrule
\end{tabular}
\label{tab:PAscoeffsghost}
\end{table}

\begin{table}[t]
\centering
\caption{Final estimates for the Taylor-series coefficients of the ghost propagator. In each line, the first uncertainty is statistical while the second is systematic, obtained from the spread of values from the different PPAs.}
\begin{tabular}{cl}
\toprule
$r_0$ & $4.17 \pm 0.01_{\mathrm{stat}}$ \\
$r_1$ & $-9.7 \pm 1.0_{\mathrm{stat}} \pm 0.5_{\mathrm{sys}}$ $\mathrm{GeV}^{-2}$ \\
$r_2$ & $33 \pm 10_{\mathrm{stat}} \pm 5_{\mathrm{sys}}$ $\mathrm{GeV}^{-4}$ \\
$r_3$ & $-110 \pm 58_{\mathrm{stat}} \pm 25_{\mathrm{sys}}$ $\mathrm{GeV}^{-4}$ \\ \bottomrule
\end{tabular}
\label{tab:finalcoeffghost}
\end{table}

As recalled above, the pattern of a pole (at about $p^2=-0.3~\text{GeV}^2$)
followed by a zero (at about $p^2=-1.0~\text{GeV}^2$) suggests that this may be the manifestation of a branch cut in $G(p^2)$.
Indeed, as described in Sec.~\ref{sec:examples},
PAs mimic the existence of a cut by accumulating a series of poles
interleaved with zeros along the cut.
On the other hand, this is clearly observed only in
very-high-order Pad\'es, since Froissart doublets may also appear
in the process,
as seen in Fig.~\ref{fig:polezerotaylor}. At the same time, if
we go to higher orders (i.e., Pad\'es with
more than 8 parameters), pole positions and zeros have --- as
already noted above --- a very large
uncertainty, of the order of 100\%, which obscures the results.
Their central values, however, do show a clear pattern of poles interleaved
with zeros. For example, in $\mathbb{P}^4_4(p^2)$ we have (apart from two
new Froissart doublets)
a pole at $-0.10~\text{GeV}^2$, followed by a zero at $-0.11~\text{GeV}^2$,
and again a pole at $-0.43~\text{GeV}^2$, and finally a zero at
$-1.3~\text{GeV}^2$.
Similarly, in $\mathbb{P}^4_5(p^2)$ one finds poles at $-0.16~\text{GeV}^2$
and $-0.84~\text{GeV}^2$, which are interleaved with two zeros,
at $-0.25~\text{GeV}^2$ and $-2.4~\text{GeV}^2$.
This pattern corroborates the idea that the pole and the zero we find in
lower-order approximants may be the first manifestation of a branch
cut, along the negative real axis of $p^2$.
We will investigate this possibility further in the next section, using
the so-called D-Log Pad\'e approximants.

\begin{figure}[t]
\centering
\includegraphics[width=0.5\textwidth]{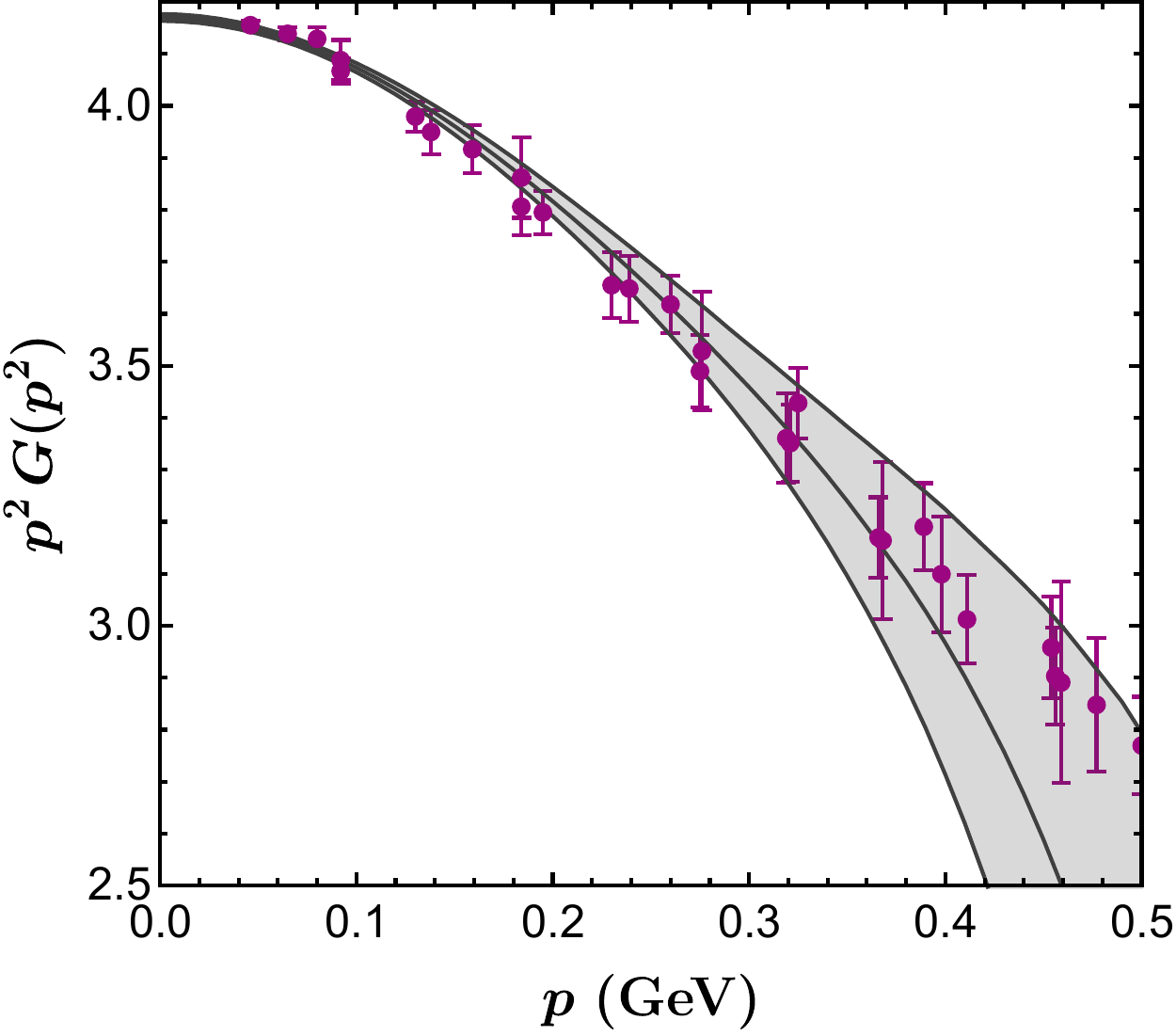}
\caption{Landau-gauge ghost-propagator Taylor series (gray) up to
third order, using the results of Table~\ref{tab:finalcoeffghost}. The
uncertainty band is the 68\% confidence level, assuming a uniform distribution
for the Taylor-series coefficients. The lattice data are shown in
purple.}
\label{fig:paseriesghost}
\end{figure}

Finally, as we did for the gluon propagator, we can extract the Taylor-series coefficients for $p^2\, G(p^2)$ around $p^2 = 0$, which can be written as
\begin{equation}
p^2 \, G(p^2) = r_0 + r_1 \, p^2 + r_2 \, p^4 + r_3 \, p^6 + r_4 \, p^8 + \cdots \; .
\label{eq:seriesghostprop}
\end{equation}
The first five coefficients $r_n$ predicted by the PPAs are shown in Table~\ref{tab:PAscoeffsghost}. One can notice that the determination of the coefficient $r_0$ is very consistent for the various Pad\'es used
and that the uncertainties increase considerably for higher orders. In order to obtain our final Taylor coefficients, we will employ all results with errors smaller than 45\%, as indicated in Table~\ref{tab:PAscoeffsghost}.
Our final estimates for the coefficients $r_n$ up to third order are
displayed in Table~\ref{tab:finalcoeffghost}, where the systematic and
statistical errors are calculated using the same procedure as before.
Assuming these values, we compare the Taylor-expanded result
for $p^2\, G(p^2)$ with the lattice data in Fig.~\ref{fig:paseriesghost}.
We see that our findings, which are model-independent, reproduce well
the behavior of the ghost-propagator data up to $p = \sqrt{p^2} = 0.5~\mathrm{GeV}$.

\subsubsection{D-Log Pad\'e approximants} \label{sec:dlog}

The results of the previous section, obtained with PPAs as fitting functions to the Landau-gauge
ghost-propagator data set, are robust and show a pole and a zero on the negative real axis of $p^2$.
However, as already stressed above, since the pole is accompanied with a zero,
one cannot exclude that this is the manifestation of a branch cut along the negative real axis.
This possibility is, in fact, corroborated by the investigation of higher-order
approximants, which do show a pattern of poles interleaved with zeros (albeit with very large
uncertainties in their positions).
We stress that, working only with the usual PAs (and PPAs), it is
quite difficult --- given the limitations imposed by the data --- to establish whether the pole
is mimicking a branch point or is, in fact, a physical pole.
In this section, we perform a first exploration of a different strategy to
extract information about a possible cut in the ghost propagator, namely we use the so-called
D-Log Pad\'e approximants~\cite{Baker1996pade}
(see also~\cite{Boito:2018rwt,Boito:2021scm} for recent applications).
The main goal of this explorative analysis is to discuss the
prospects, the advantages, and the limitations of D-Log PAs when applied to the Landau-gauge
ghost-propagator data.

The idea behind D-Log PAs is to manipulate a function having a cut in the
complex plane into a form that can be better approximated by PAs, and later
unfold the procedure.
To this end, let us suppose we are interested in the following function
\begin{equation}
f(z) = A(z) \, \dfrac{1}{(\mu - z)^\gamma} + B(z) \; ,
\label{eq:funcdlog}
\end{equation}
where $A(z)$ and $B(z)$ have a simple structure and are analytic at $z=\mu$,
and $\gamma$ is not necessarily an integer --- so that $f(z)$ can have a cut with branch point $z=\mu$.
Instead of working directly with $f(z)$, we use a
related function $F(z)$ given by
\begin{equation}
F(z) = \dfrac{\mathrm{d}}{\mathrm{d} z} \ln{f(z)} \approx \dfrac{\gamma}{(\mu - z)} \; ,
\label{eq:Fdlog}
\end{equation}
where on the right-hand side we assume $z\approx \mu$.
Clearly, the function $F(z)$ is, in this case, meromorphic and the
usual convergence theorems apply.
We stress that the exponent of the cut is now the residue of the simple pole
of $F(z)$, which can, in principle, be determined in an unbiased
way using Pad\'e approximants --- denoted in this context by $\bar{P}^M_N(z)$
--- to $F(z)$.
Then, by unfolding this procedure, the original function $f(z)$
is approximated by a non-rational function, the so-called D-Log Pad\'e approximant, given by
\begin{equation}
\mathrm{Dlog}_N^M(z) = f_{\mathrm{norm}}(0) \, \exp{\left\{\int \mathrm{d}z \,\, \bar{P}_N^M(z) \right\}} \; ,
\label{eq:dlogpa}
\end{equation}
where the constant $f_{\mathrm{norm}}(0)$ has to be adjusted to reproduce the function $f(z)$
at zero, since the constant value $f(0)$ --- in the Taylor expansion of $f(z)$
--- is lost due to the derivative in Eq.~(\ref{eq:Fdlog}).

In view of the fact that in this work we use the PAs as fitting functions to lattice data,
the application of D-Log PAs to our problem requires
some additional steps, in comparison with the usual PAs. Indeed, we
need (lattice) data for the function $F(p^2)$, i.e., the derivative
of the logarithm of the ghost propagator $G(p^2)$.
Clearly, these data for $F(p^2)$ can
be obtained from the propagator data, by first evaluating
the logarithm and, then, by taking the numerical
derivative, using finite differences. However, the application of the standard
formulas for the derivative requires equally-spaced data points, a
property that is not satisfied by lattice data for the ghost propagator.
Thus, in order to
circumvent this problem, we need first to interpolate between the data points,
as a means to generate a new set of data on a uniform grid. We do this using a
linear interpolation, which should be sufficient for our
purposes.
More specifically, after taking the logarithm of the central values
of the ghost-propagator data, we interpolate them and generate
equally-spaced data points, with a separation of
$\Delta p^2= 0.035~\text{GeV}^2$. We also propagate
the errors accordingly.
One should note, however, that the linear interpolation
introduces correlations between some of the new data points, since they may share information of the original ghost-propagator data.
We calculate these correlations using standard
methods~\cite{Boito:2020xli,Keshavarzi:2018mgv} and take them into account
carefully, in all of the subsequent results.
Next, the derivative at the $i$-th point of
this uniform grid, $p_i$, can be calculated using, for example,
the usual first-order formula
\begin{equation}
f'_{(1)}(p^2_i) \approx \dfrac{f(p^2_{i+1})-f(p^2_{i-1})}{2 \, \Delta p^2} \; .
\end{equation}
This procedure introduces, in principle, another source of
correlations in the data, since the derivative obtained at nearby
points may share information from the same underlying data point.
This problem can be circumvented exactly by skipping the appropriate number of points,
in such a way that each derivative depends
on different entries of the original data set.
(For example, in the case of the above first-order formula, this is
achieved when the derivative is evaluated at every third point.)
Another possibility, which we will use here, is to compute the first-order
derivative at every second point, calculating all the correlations involved in
the process, which again can be done in a rather straightforward manner.
The advantage of this procedure is that the final data set contains
enough points, while keeping the correlations relatively mild.
(On the other hand, when computing the derivative at every point
$p_i^2$, the correlations become too strong to allow for a meaningful
fit.) We have also considered
higher-order numerical derivatives. The number
of data points, however, decreases rapidly when the order is
larger than two (if one wants to reduce correlations in the data
by skipping some of the points) and, for this reason, we will
only consider below first-order derivatives.

\begin{figure}[t]
\centering
\subfloat[]{\label{fig:numder1}{\includegraphics[width=0.49\textwidth]{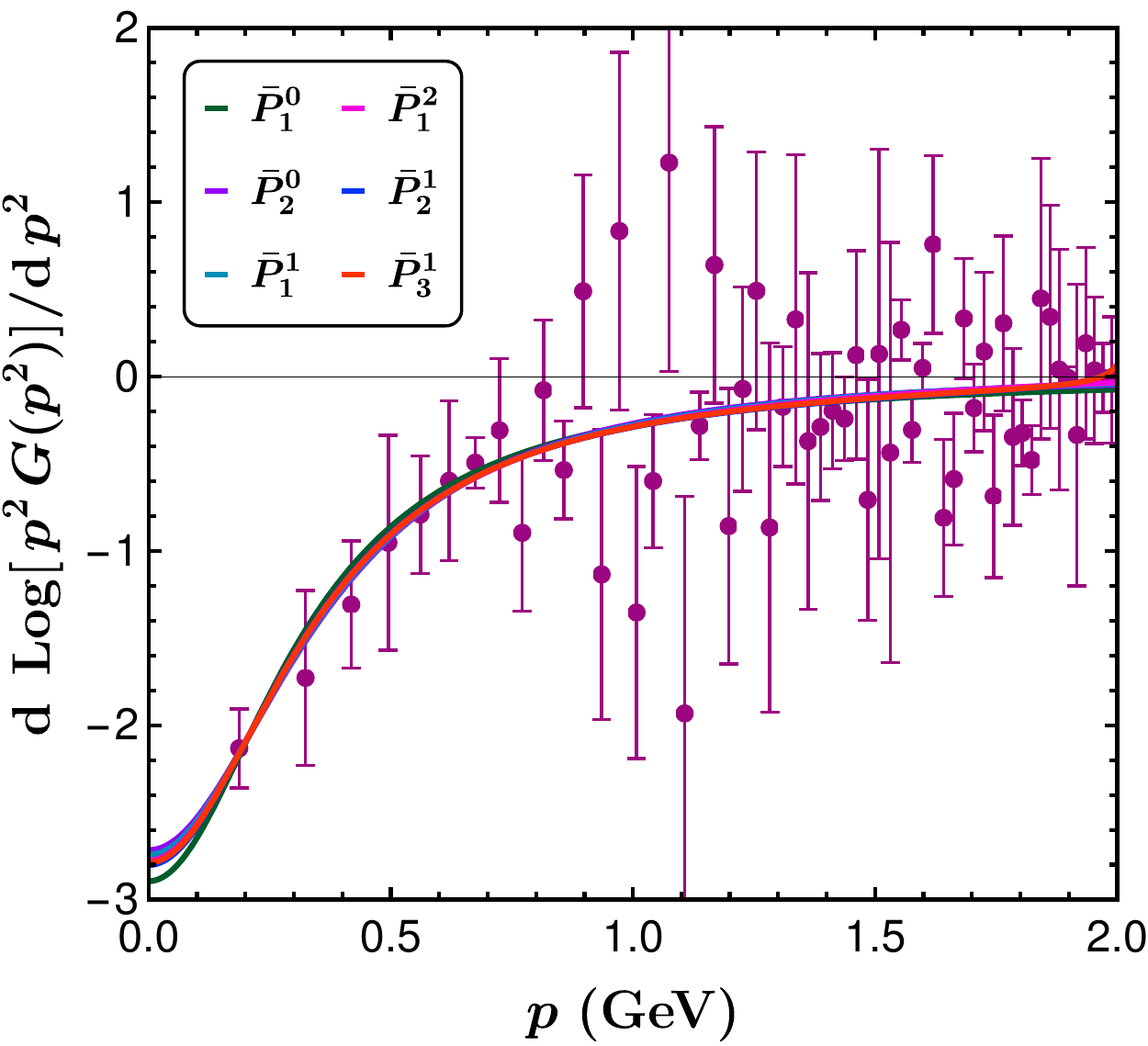}}}\hfill
\subfloat[]{\label{fig:dlogdataghost}{\includegraphics[width=0.47\textwidth]{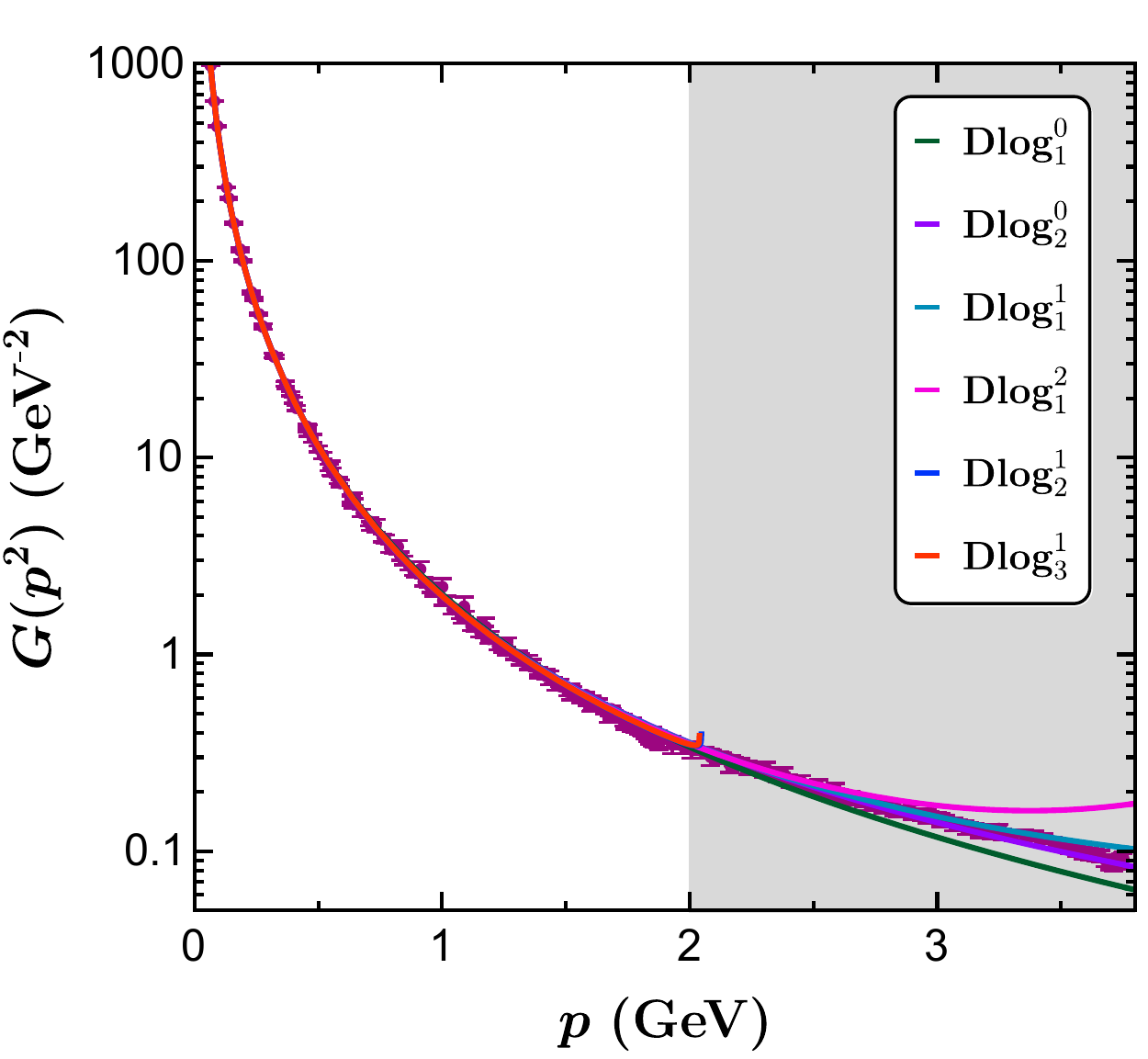}}}
\caption{(a) First-order numerical derivative of the logarithm of the ghost-propagator data together with the Padés $\bar{P}_M^N(p^2)$ used to build the D-Log PAs. (b) The lattice data for the ghost propagator and the D-Log Padé approximants built from the first-order numerical derivative of the logarithm of the data.
The shaded region --- corresponding to $p = \sqrt{p^2} > 2.0~\text{GeV}$ ---
is excluded from the fit.}
\end{figure}

As a check, we also did the same analysis described above using the bootstrap method
with 5000 samples (and verified that results do not change when using 2500 samples).
In all cases we find very good agreement between the data sets produced by the two
analyses.
In Fig.~\ref{fig:numder1}, we show the results for the derivative of the
logarithm of the ghost-propagator data, evaluated numerically
at every second point, computing
all the ensuing correlations to the data.
It is evident that the uncertainties in the final
data set are rather large --- of the order of 100\% in some
cases --- with significant statistical fluctuations (as one could expect).
We, nevertheless, carry on with our analysis in order to understand the applicability
and the limitations of the D-Log method for the data set considered here.

We first perform the minimization
to the (generated) data set, using the PAs $\bar{P}_N^M(p^2)$ as
fitting functions. The result is shown in Fig.~\ref{fig:numder1}.
We again use diagonal fits, i.e., as said above, we consider in
the fit quality only the diagonal elements of the covariance matrix.
In this case, for the fit, we select data in the interval
$p \leq 2.0~\text{GeV}$, since at larger momenta the errors and fluctuations
are very large. Moreover, as clearly seen from Fig.~\ref{fig:numder1}, the data
at $p^2 = 2.0~\text{GeV}$ are already described by the perturbative $1/p^2$
behavior. After the minimization, Eq.~(\ref{eq:dlogpa}) is employed to build
the D-Log Pad\'es to the ghost-propagator data.
We performed fits using the D-Log sequences
$\mathrm{Dlog}_{N+1}^N$,
$\mathrm{Dlog}_{N+2}^N$, $\mathrm{Dlog}_N^N$ and $\mathrm{Dlog}_N^{N+1}$.
As before, the maximum value of $N$ that can be considered is limited
by the quality of the fitted data. In particular, in this case, the
maximum number of parameters that lead to meaningful results is
only 5, which is a consequence of the large
errors and fluctuations of the data set.
We find that these approximants have a branch cut of the form
\beq
\frac{1}{(p_c-p^2)^\gamma} \; .
\eeq
The branch point $p_c$ and the multiplicity
$\gamma$ of the cut obtained from different D-Log approximants
are reported in Table~\ref{tab:Dlogcutghost}, together with the corresponding
$Q^2/\mathrm{dof}$ value. One can observe that
the approximants that pass all reliability tests have indeed a cut on the negative real
axis, with a branch point not too far from the pole at $-0.30~\text{GeV}^2$,
determined in the
previous section. Our final value for the branch-point position is
\begin{equation}\label{eq:pcfinal}
p_c = (-0.12\pm 0.08_{\rm stat}\pm 0.02_{\rm sys})~\mathrm{GeV}^2 \; .
\end{equation}
One should stress, however, that our systematic uncertainty
for $p_c$ may be underestimated, since the evaluation
of the numerical derivative inevitably introduces an additional
source of error.
At the same time, the exponent $\gamma$ is clearly not compatible
with one, which reinforces that the function has indeed a cut (in the negative
real axis) and not a pole.
Our results for the D-Log Padés, together with the original
lattice data, are shown in Fig.~\ref{fig:dlogdataghost},
where it is possible to observe a reasonably good agreement, even beyond the fit region.

\begin{table}[t]
\centering
\caption{Branch point $p_c$ and the multiplicity $\gamma$,
alongside the corresponding $Q^2/\mathrm{dof}$ value, from fits of
D-Log Padé approximants to the Landau-gauge ghost-propagator lattice data.}
\begin{tabular}{cccc}
\toprule
D-Log PA & $p_c$ \small{$(\mathrm{GeV}^2)$} & $\gamma$ & $Q^2/\mathrm{dof}$ \\ \midrule
$\mathrm{Dlog}_1^0$ & $-0.11 \pm 0.03$ & $0.31 \pm 0.05$ & 0.81 \\
$\mathrm{Dlog}_2^0$ & $-0.14 \pm 0.07$ & $0.4 \pm 0.2$ & 0.81 \\
$\mathrm{Dlog}_1^1$ & $-0.13 \pm 0.05$ & $0.4 \pm 0.1$ & 0.81 \\
$\mathrm{Dlog}_1^2$ & $-0.12 \pm 0.09$ & $0.3 \pm 0.2$ & 0.83 \\
$\mathrm{Dlog}_2^1$ & $-0.12 \pm 0.04$ & $0.3 \pm 0.2$ & 0.83 \\
$\mathrm{Dlog}_3^1$ & $-0.12 \pm 0.08$ & $0.3 \pm 0.2$ & 0.85 \\
\bottomrule
\label{tab:Dlogcutghost}
\end{tabular}
\end{table}

\begin{figure}[!t]
\centering
\includegraphics[width=0.49\textwidth]{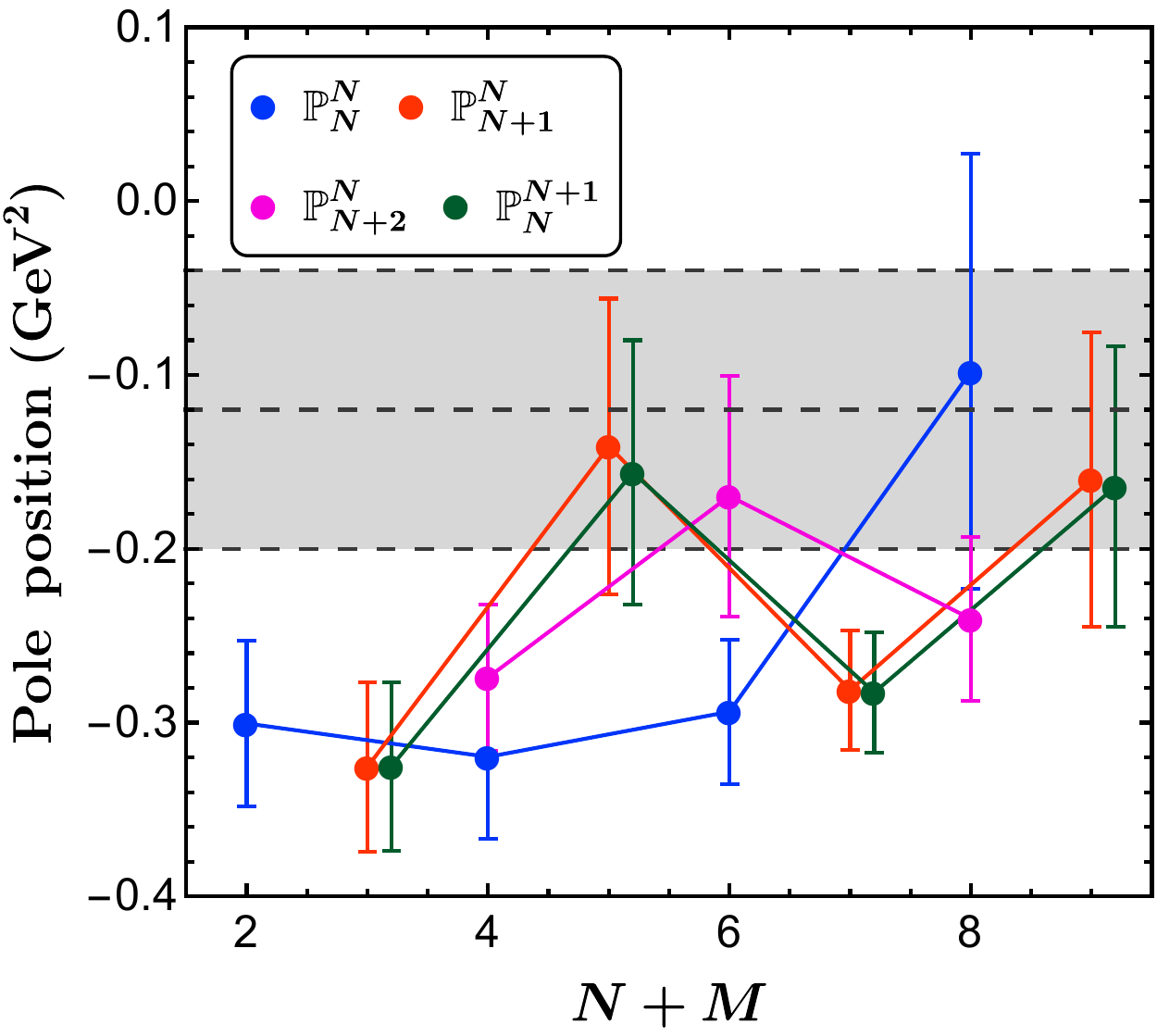}
\caption{Comparison of the largest (real and nonzero) pole position from four partial Padés sequences with the branch point position obtained form the D-Log approximants, Eq.~(\ref{eq:pcfinal}), shown as the gray band. At higher-orders the position of the first pole tends to approach the D-Log branch point, indicating that the PPAs are mimicking the cut. We note that the uncertainty on the gray band may be underestimated,
since the application of the D-Log Pad\'es requires the calculation of numerical derivatives from the original data set.}
\label{fig:poleconvghost}
\end{figure}

Let us now compare the results from the D-Log approximants
with the results from the PPAs, described in the previous section.
It is evident that the analysis using D-Log approximants
does corroborate the existence of a branch cut on the negative real axis of $p^2$.
Then, if the cut is indeed present, one would expect that the pole --- found
with the PPAs --- should indicate the position of the branch point.
As a matter of fact, the final results we obtain for $p_c$ and for the pole
extracted from the PPAs [see Eqs.~(\ref{eq:pcfinal}) and~(\ref{eq:pologhostfinal})]
are compatible within $1.7\sigma$.
Furthermore, we observe that, for higher-order PPAs,
the position of the pole tends to larger values, i.e.\ to $-0.10~\text{GeV}^2$ and
$-0.16~\text{GeV}^2$ respectively when considering $P^4_4(p^2)$ and $P^4_5(p^2)$,
thus approaching the value of $p_c$ from the
D-Log approximants, although with larger errors.
This can be clearly seen in Fig.~\ref{fig:poleconvghost}, where we compare $p_c$ from
Eqs.~(\ref{eq:pcfinal}) with the largest (nonzero) pole positions from different
PPA sequences, with up to 10 parameters.
It is evident that higher-order PPAs show a pole position in better agreement
with the location of $p_c$ obtained from the D-Log approximants. This may indicate a possible
convergence between the two methods.

In conclusion, the results of this section support, apart from the pole at
$p^2=0$, the existence of a branch cut along negative values of $p^2$.
Our best estimator for the branch point $p_c$ is obtained from the D-Log approximants,
and this value is in reasonably good agreement with the results from the PPAs.

\section{Conclusions}
\label{sec:conclusions}

In this work we have used rational approximants in order to study the analytic
structure of the IR Landau-gauge gluon and ghost propagators from a
model-independent
point of view. With this method, we do not rely on any specific theoretical
description of the propagators in the IR. Instead, the main analytic features
are obtained from a systematic use of Pad\'e approximants (and
also of partial Pad\'e and D-Log
Pad\'e approximants) as fitting functions to the lattice
data for the $SU(2)$ gluon and ghost propagators.
A closely related work, applying Pad\'e approximants to the $SU(3)$ propagators,
may be found in Refs.~\cite{Falcao:2020vyr,Oliveira:2021job}.
The results we present here have more robust uncertainties, since we have performed, in all cases, a careful error propagation,
including all correlations in the data set as well as in the fit parameters.
Apart from the statistical error, we have also estimated a
theoretical systematic error, inherent to the method.
Another important difference with respect to
Refs.~\cite{Falcao:2020vyr,Oliveira:2021job}
is that we have exploited variants of the usual Pad\'e approximants, namely
the partial Pad\'es and the D-Log Pad\'es, which were particularly useful in
the study of the ghost propagator.

In the case of the gluon propagator, the Pad\'e approximants show distinct evidence
for a pair of complex poles located at
\begin{equation}
p^2 = [(-0.37 \pm 0.09) \pm i \,(0.66 \pm 0.04)] \,\, \mathrm{GeV}^2 \; ,
\end{equation}
where we added in quadrature the statistical and systematic errors of Eq.~(\ref{eq:polegluonfinal}).
This result is of course in good
agreement with the analysis presented in Ref.~\cite{Cucchieri:2011ig}
(see their Table IV) for the same set of lattice data.
It is also remarkable that the pole position turns out very similar to
those found in the $SU(3)$ gluon
propagator,
namely $p^2 = -0.28(6)\pm i\,0.4(1)~\text{GeV}^2$ or $p^2 = -0.19(4)\pm i\,
0.4(1)~\text{GeV}^2$ --- depending on the method used in the
analysis of Ref.~\cite{Falcao:2020vyr} ---
and $p^2= -0.30(7) \pm i \,0.49(3)~\text{GeV}^2$ from
Ref.~\cite{Binosi:2019ecz}.\footnote{We stress, however, that in these results the errors are solely systematic.}
At the same time, we find evidence for a zero in the gluon propagator,
located at $p^2_{\rm zero}=(-2.9 \pm 1.0)~\text{GeV}^2$.

As for the ghost propagator, the analysis with PAs reveals
that the approximants require a pole at the origin.
This input has, therefore, been used to build the so-called partial
Pad\'e approximants, which provide evidence for a pole --- along
the negative real axis --- followed by a zero.
At the same time, higher-order PPAs also show additional intercalated
poles and zeros. As stressed in the text, this result is an indication
that the approximants may be mimicking a branch cut of the function by
accumulating poles (and zeros) along the cut~\cite{Baker1996pade}.
In order to test this hypothesis, we considered in
Sec.~\ref{sec:dlog} the so-called D-Log Pad\'e approximants, which
are particularly useful to study functions with a branch cut. Indeed, we
found evidence for a cut with a branch point located at
$p^2 =( -0.12 \pm 0.08)~\text{GeV}^2$,
in excellent agreement with
Refs.~\cite{Falcao:2020vyr,Oliveira:2021job}, which report
a cut starting around $p^2 \sim -0.1~\text{GeV}^2$. One should stress,
however, that the quoted uncertainty in our branch-point location may be
underestimated, since the application of D-Log approximants requires the
calculation of numerical derivatives of the data for $G(p^2)$, which inevitably
introduces additional errors.
The value of this branch point is, nevertheless, in reasonable
agreement with the pole position
$p^2_{\rm pole} = (-0.30 \pm 0.07)~\text{GeV}^2$ found
using the PPAs. This corroborates, in
our opinion, that the poles and zeros displayed by the PPAs
for $p^2 < 0$
are, in fact, mimicking a cut along the negative real axis of $p^2$.
We also note that these results go along with the analysis presented
in Ref.~\cite{Cucchieri:2008fc}, where the ghost dressing function
$p^2 G(p^2)$ is well described by a fitting form displaying a logarithmic
cut in the negative real axis of $p^2$, with a branch point very close
to zero.

We note that a similar analysis with PPAs and D-Log PAs in the
gluon-propagator case is inconclusive regarding the presence of a
branch cut. This is due to the limitations imposed by the lower quality
of the data and by the richer structure of the function, which did not enable
us to consider --- in a controlled way --- approximants of sufficiently high
order for the gluon propagator.
Finally, the PAs (for gluon) and PPAs (for ghost) also allowed for a model-independent determination of the
first few coefficients of the Taylor-series expansion of $D(p^2)$
and $G(p^2)$ around $p^2=0$, which serve as an
additional constraint to the theoretical description of
gluon and ghost propagators at small momenta.

As said above, our analysis strongly supports the existence of a pair of
complex-conjugate poles in the analytic structure of the gluon propagator,
in agreement with Refs.~\cite{Binosi:2019ecz,Falcao:2020vyr,Oliveira:2021job}
and with the so-called Gribov-Zwanziger scenario~\cite{GRIBOV19781,
Vandersickel:2011zc,Vandersickel:2012tz}.
Furthermore, the Pad\'e analysis strongly disfavors the possibility of real
poles in the gluon sector, as considered, e.g., in Ref.~\cite{Li:2019hyv}.
Note also that the presence of complex poles in the gluon propagator
has been recently criticized in Ref.~\cite{Horak:2022myj}, since
this would imply the violation of the K\"all\'en-Lehmann representation for
the ghost propagator (as already recalled in the Introduction). The authors
claim that no such violation has yet been observed, quoting
Refs.~\cite{Binosi:2019ecz,Dudal:2019gvn,Fischer:2020xnb,Falcao:2020vyr}.
We note, however, that the cited references do not make such a claim.
In fact, Ref.~\cite{Falcao:2020vyr} does not make any statement about the
existence of the K\"all\'en-Lehmann representation for the ghost propagator.
At the same time, in Refs.~\cite{Binosi:2019ecz,Dudal:2019gvn}
the existence of this representation is the starting point
of the analyses, i.e.\ these two works have verified that numerical data
are well described by using such a representation. Of course, as discussed
below, this does not constitute a proof of its validity.
Lastly, Ref.~\cite{Fischer:2020xnb} does not appear to be conclusive
on the matter since, as stressed by the authors in their conclusions, the
results strongly depend on the details of the chosen model for the
gluon-ghost vertex.

As a final remark, we would like to reinforce that our analysis
clearly highlights the limitations imposed by the data sets: fits with
too many parameters do not add new information, since they lead to
quantities with huge errors and parameters that are 100\% correlated.
In fact, we know --- for example from
Refs.~\cite{Cucchieri:2011ig,Cucchieri:2016jwg} --- that a very good
description of lattice data for Landau-gauge gluon and ghost propagators
may be easily achieved with reasonably simple functions, depending
on just a few parameters.
For this reason, reproducing lattice data via complicated analytic expressions,
depending on a large set of parameters (such as those considered in
Refs.~\cite{Binosi:2019ecz,Dudal:2019gvn,Falcao:2020vyr,Oliveira:2021job}) is
very likely to work, but it is not, by itself, a sufficient criterion to
validate the considered fitting functions.
As a consequence, the extraction of information from this type of analysis,
without careful consideration of uncertainties and correlations,
should be taken with caution.

\section*{Acknowledgements}

We thank Pere Masjuan for insightful comments on the numerical precision
required for higher-order approximants, which motivated the inclusion of
Appendix~\ref{app:Precision}.
DB's work was supported by the S\~ao Paulo Research Foundation (FAPESP) grant
No.\ 2021/06756-6 and CNPq grant No.\ 308979/2021-4.
AC and TM acknowledge partial support from FAPESP and CNPq.
The work of CYL was financed by FAPESP grants No.\ 2020/15532-1 and No.\ 2022/02328-2 and CNPq grant
No.\ 140497/2021-8.

\appendix

\section{Fits to highly correlated data}
\label{sec:AppCorrData}

When dealing with strongly correlated data, standard $\chi^2$ fits can lead to biased or unreliable results. This stems from the tiny eigenvalues of the covariance matrix, which must be inverted in the usual $\chi^2$-fit construction. An alternate procedure is to employ a method, which we will call ``diagonal fits'', where only the diagonal elements of the covariance matrix are considered in the fit quality, named, in this case, $Q^2$~\cite{Boito:2011qt}.

The fit quality $Q^2$ is obtained from the diagonal elements of the data covariance matrix, $C$. We denote by $C_0$ the diagonal matrix obtained from $C$ neglecting off-diagonal terms. Both $C$ and $C_0$ are symmetric and positive-definite matrices. We consider then the fit quality
\begin{equation}
Q^2 = [d_i - f_i(\vec{a})](C_0^{-1})_{ij}[d_j - f_j(\vec{a})] \; ,
\end{equation}
where $d_i$ are elements of binned data set and $f_i(\vec{a})$ is the fitting function
(the PAs in this work) with a vector of parameters $\vec{a}$ that describes the data.
(In the above equation repeated indices are summed over.)
Since off-diagonal correlations are not being considered in the fit quality, we lose the statistical interpretation of the $\chi^2$ and it does not make sense to judge the fit quality in absolute terms; in particular, the usual $p$-value cannot be used.

With this setup, the
parameters $\vec{a}$ are determined by minimizing $Q^2$, i.e.\ by
solving the system of equations
\begin{equation}\label{eq:Q2sol}
\dfrac{\partial Q^2}{\partial a_\alpha} = -2 \, \dfrac{\partial f_i(\vec{a})}{\partial a_\alpha} \, (C_0^{-1})_{ij} [d_j - f_j(\vec{a})] = 0 \; ,
\end{equation}
which can be done numerically or, in a linear case,
for example, even analytically.

We now turn to the evaluation of the errors on the parameters
$a_\alpha$.
By varying the parameters and the data by $\delta a_\alpha$ and $\delta d_i$, respectively, in Eq.~(\ref{eq:Q2sol}) we can relate $\delta a_\alpha$ and $\delta d_i$ as
\begin{equation}
\delta a_\alpha = A_{\alpha\beta}^{-1} \, \dfrac{\partial f_i (\vec{a})}{\partial a_\beta} \, (C_0^{-1})_{ij} \, \delta d_j \; ,
\end{equation}
where we considered that the fit is good, i.e., the difference $(d_i - f_i(\vec{a}))$ is small and can be ignored (which kills the term with the second derivative), and we defined the matrix
\begin{equation}
A_{\alpha\beta} = \dfrac{\partial f_i(\vec{a})}{\partial a_\alpha} \, (C_0^{-1})_{ij} \, \dfrac{\partial f_j(\vec{a})}{\partial a_\beta} \; .
\end{equation}
The fit-parameter covariance matrix $\langle \delta a_\alpha \, \delta a_\beta \rangle$ is then given by
\begin{equation}\label{eq:parmcovQ2}
\langle \delta a_\alpha \, \delta a_\beta \rangle = (A^{-1})_{\alpha\gamma} \, \dfrac{\partial f_i(\vec{a})}{\partial a_\gamma} \, (C_0^{-1})_{ik} \, C_{kl} \,(C_0^{-1})_{lj} \, \dfrac{\partial f_j(\vec{a})}{\partial a_\sigma} \, (A^{-1})^T_{\sigma\beta} \; ,
\end{equation}
where $C_{ij} = \langle \delta d_i \, \delta d_j \rangle$ is the complete data covariance matrix, including all off-diagonal terms. The equation above is an estimative for the full covariance matrix of the parameters of the fit $\vec{a}$.
We stress that, if the fit is good enough to disconsider terms proportional to
$(d_i - f_i(\vec{a}))$ and when the matrix $C_{0,ij}$ is replaced by the full
matrix $C_{ij}$\, this method gives a covariance matrix equivalent to the one obtained
from the Hessian of the usual $\chi^2$ procedure. The advantage of Eq.~(\ref{eq:parmcovQ2}) is that the full data covariance matrix $C$ is never inverted in the procedure, avoiding the issues related to its small eigenvalues. The effects of the data correlation are, nevertheless, fully taken into account in the error propagation, as shown in Eq.~(\ref{eq:parmcovQ2}).

\section{Numerical precision in higher-order Pad\'e approximants}
\label{app:Precision}

\begin{figure}[H]
\centering
\includegraphics[width=0.4\textwidth]{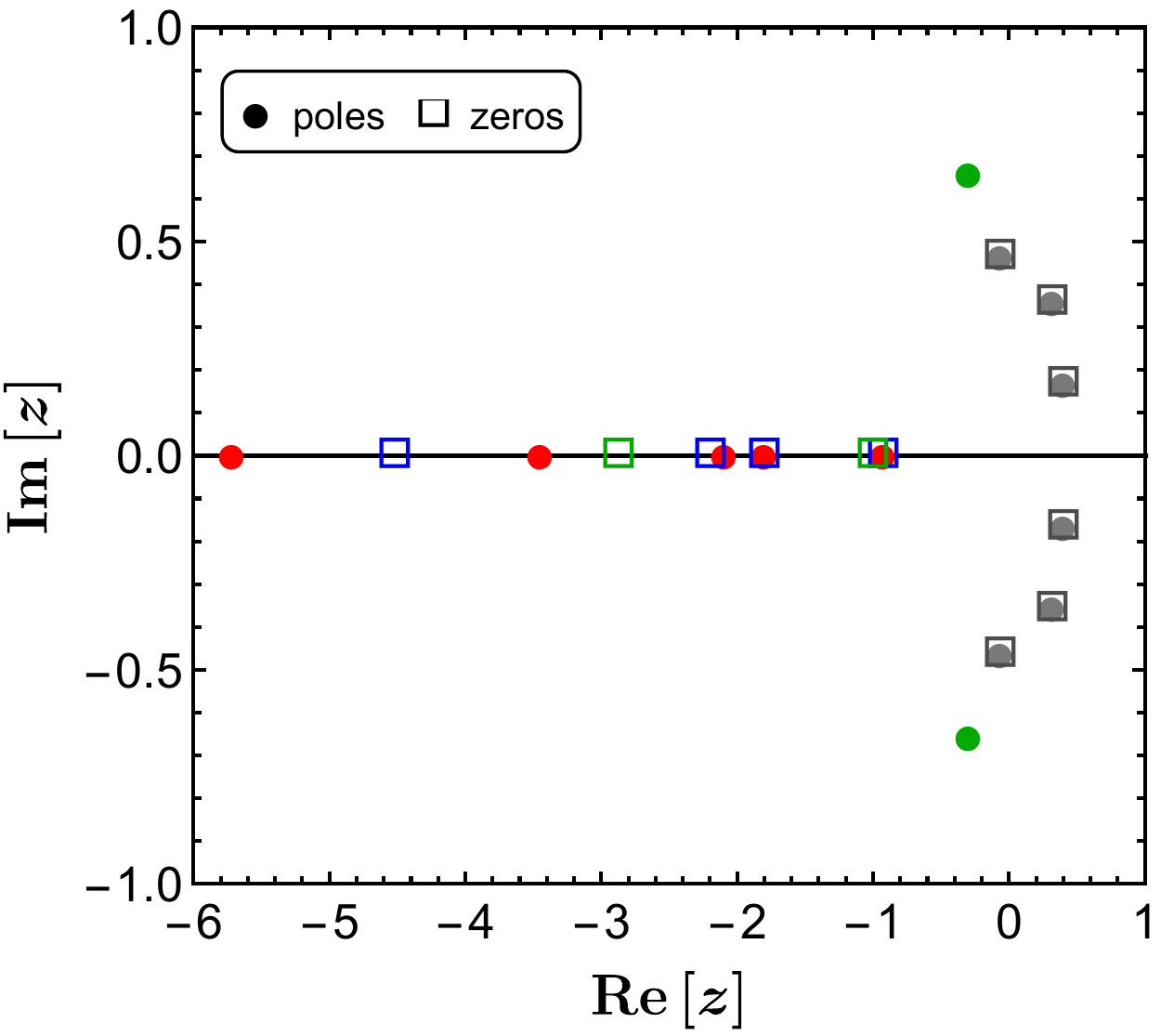}
\caption{Results obtained with insufficient numerical precision
(see discussion in the text) for the poles (filled circles) and zeros
(empty squares) of $P_{15}^{14}(z)$, built to the Taylor series of $g(z)$
defined in Eq.~(\ref{eq:logpole}). The poles and zeros
that can be identified with genuine poles of $g(z)$ are shown
in green, while artifacts are shown in red (poles) and blue (zeros). It is
possible to see a number of spurious complex poles and zeros (shown in dark gray),
resembling Froissart doublets, that are the result of round-off errors.
The correct result can be seen in Fig.~\ref{fig:polezerotaylor}.}
\label{fig:precisionprob}
\end{figure}

As mentioned in Sec.~\ref{sec:examples}, when building higher-order Pad\'e
approximants, it is very important to work with sufficiently high numerical
accuracy or, otherwise, the results can be contaminated with spurious poles
and zeros.
In order to illustrate this, we consider again the use of
the genuine Pad\'es to approximate the function $g(z)$ of
Eq.~(\ref{eq:logpole}) for the case $P^{14}_{15}(z)$, which corresponds to
30 parameters.
As said before, we do not expect any complex
Froissart doublets to appear in this case.
This is because $\log(z+2)$ is a Stieltjes
function, which generates only real poles in the PAs~\cite{Baker1975essentials,Baker1996pade},
while the remainder of $g(z)$ is a rational meromorphic function, very easily reproduced by
the Pad\'es. However, if insufficient numerical precision is used,
the solution to the system of 30 equations
can lead to wrong results, in which some of the poles
move from the real axis into the complex plane, resembling Froissart
doublets (see Sec.~\ref{sec:examples}).
These numerical artifacts may appear in a semi-circular pattern~\cite{ellipse},
as can be clearly seen in Fig.~\ref{fig:precisionprob}, and are similar to the doublets that appear due to the presence of random noise in the input Taylor coefficients~\cite{ellipse,MasjuanQueralt:2010hav}.
Let us mention that this feature is also visible in the results of
Ref.~\cite{Falcao:2020vyr}, both in the analysis of their toy data and of
the lattice data.

In order to circumvent this problem, in the results of Sec.~\ref{sec:examples},
we employed a precision of 30 decimal places for our analysis of the
function $g(z)$. In particular, the approximants shown in
Fig.~\ref{fig:toydata} were obtained applying the built-in function
\texttt{Pad\'eApproximant} of {\tt Wolfram Mathematica}, imposing this
precision. On the contrary, in the above (wrong) solution the same
function was used with default precision.
Of course, in the presence of spurious pairs of complex zeros
and poles generated by round-off errors, it is more difficult to
identify the cut of the function $g(z)$. This occurs since some of the poles
and zeros that would appear along the cut are now
displaced into the complex plane.

\vspace{-0.2cm}
\bibliographystyle{jhep}
\bibliography{References}

\end{document}